\documentclass[runningheads,a4paper]{llncs}

\usepackage{amssymb}
\usepackage{graphicx}
\usepackage[czech,english]{babel}

\usepackage{url}
\urldef{\mailsa}\path|{alfred.hofmann, ursula.barth, ingrid.haas, frank.holzwarth,|
\urldef{\mailsb}\path|anna.kramer, leonie.kunz, christine.reiss, nicole.sator,|
\urldef{\mailsc}\path|erika.siebert-cole, peter.strasser, lncs}@springer.com|    
\newcommand{\keywords}[1]{\par\addvspace\baselineskip
\noindent\keywordname\enspace\ignorespaces#1}
\usepackage{epsfig}
\usepackage{latexsym}
\usepackage{amsmath}
\usepackage{amssymb}
\usepackage{latexsym}
\usepackage{times}
\usepackage{epsfig}
\usepackage{subfigure}
\usepackage{graphicx}
\usepackage{geometry}
\usepackage{rotating}
\newcommand{\Vspace}[1]{ }
	
\usepackage{t1enc}
\usepackage{amsmath}
\usepackage{amsfonts}
\newcommand{\kth}[0]{$k^\text{th}$ }
\newcommand{\mbf}[0]{{\sc mbf}}
\newcommand{\av}[0]{{\sc av}}

\usepackage{graphicx}

\begin{document}

\mainmatter  

\title{Efficient Minimization of Higher Order Submodular Functions using Monotonic Boolean Functions}

\titlerunning{Efficient Minimization of Higher Order Submodular Functions using Monotonic Boolean Functions}

\author{Srikumar~Ramalingam$^1$ \quad Chris~Russell$^{2\&3}$ \quad \v Lubor Ladick\'{y}$^4$ \quad Philip~H.S.~Torr$^5$}
\institute{
$^1$University of Utah, USA\\
$^2$Alan Turing Institute, UK \\
$^3$University of Edinburgh, UK \\
$^4$ETH Zurich, Switzerland \\
$^5$University of Oxford, Oxford, UK \\
}
\authorrunning{Srikumar Ramalingam, Chris Russell, \v Lubor Ladick\'{y}, and Philip~H.S.~Torr}
\maketitle

\begin{abstract}
Submodular function minimization is a key problem in a wide variety of applications 
in machine learning, economics, game theory, computer vision, and many others. The 
general solver has a complexity of $O(n^3 \log^2 n . E +n^4 {\log}^{O(1)} n)$ 
where $E$ is the time required to evaluate the function and $n$ is the number of 
variables \cite{Lee2015}. On the other hand, many computer vision and machine learning 
problems are defined over special subclasses of submodular functions that can be written as the sum of many submodular cost functions defined over cliques containing few variables. In such functions, the pseudo-Boolean (or polynomial) representation \cite{BorosH02} of these subclasses are of degree (or order, or clique size) $k$ where $k \ll n$. In this work, we develop efficient algorithms for the minimization of this useful subclass of submodular functions. To do this, we define novel mapping that transform submodular functions of order $k$ into quadratic ones. The underlying idea is to use auxiliary variables to model the higher order terms and the transformation is found using a carefully constructed linear program. In particular, we model the auxiliary variables as monotonic Boolean functions, allowing us to obtain a compact transformation using as few auxiliary variables as possible. The transformed quadratic function can be 
efficiently minimized using the standard max-flow algorithm with a time complexity 
of $O((n+m)^3)$ where $m$ is the total number of auxiliary variables involved in 
transforming all the higher order terms to quadratic ones. Specifically, we show 
that our approach for fourth order function requires only $2$ auxiliary variables 
in contrast to $30$ or more variables used in existing approaches. In the general 
case, we give an upper bound for the number or auxiliary variables required to 
transform a function of order $k$ using Dedekind number, which is substantially 
lower than the existing bound of $2^{2^k}$.

\keywords{submodular functions, quadratic pseudo-Boolean functions, monotonic 
Boolean functions, Dedekind number, max-flow/mincut algorithm}
\end{abstract}

\section{Introduction}
\Vspace{-0.2cm} Many optimization problems in several domains such as
operations research, computer vision, machine learning, and
computational biology involve submodular function
minimization. Submodular functions (See Definition~\ref{def.sbf}) are
discrete analogues of convex functions \cite{lovasz83}. Examples of
such functions include cut capacity functions, matroid rank functions
and entropy functions. Submodular function minimization techniques may
be broadly classified into two categories: algorithms for general 
submodular functions and efficient and customized algorithms for 
subclasses of submodular functions. This paper falls under the 
second category.

\paragraph{General solvers:} The role of submodular functions in
optimization was first discovered by Edmonds when he gave several
important results on the related poly-matroids
\cite{Edmonds69}. Gr{\"o}tschel, Lov{\'a}sz, and Schrijver first gave a
polynomial-time algorithm for minimization of submodular function
using ellipsoid method \cite{GrotschelLS81}.
Recently several combinatorial and strongly polynomial algorithms
\cite{Fleischer01apush-relabel,Iwata00afully,Iwata01acombinatorial,Schrijver2000,orlin09}
have been developed based on the work of Cunningham
\cite{cunningham85}. The current best strongly polynomial algorithm 
for minimizing general submodular functions~\cite{Lee2015} has a 
run-time complexity of $O(n^3 \log^2 n . E +n^4 {\log}^{O(1)} n)$, 
where $E$ is the time taken to evaluate the function, and $n$ is the 
number of variables. Weakly polynomial time algorithms with a smaller 
dependence on $n$ also exist. For example, Lee et al.~\cite{Lee2015} 
shows a method with a run-time complexity of $O(n^2 {\log}~nM . E + 
n^3 {\log}^{O(1)} nM)$, where $M$ is the maximum absolute value of the 
function values. 

\paragraph{Specialized solvers:} 
Higher order submodular functions are useful in modeling many computer 
vision and machine learning problems \cite{kohlicvpr07,LanRHB06,ishikawa2011}. 
Such problems typically involve millions of pixels making the use of general
solvers highly infeasible. Further, each pixel may take multiple
discrete values and the conversion of such a problem to a Boolean one
introduces further variables. On the other hand, the
cost functions for many such optimization algorithms belong to a 
small subclass of submodular functions. The goal of this paper is to
provide an efficient approach for minimizing these subclasses of
submodular functions using a max-flow algorithm.  \Vspace{-0.2cm}

\paragraph{Notations:}
Let $\mathbb B$ denote the Boolean set $\{0,1\}$ and $\mathbb R$ the set of reals. Let the vector ${\bf x}=(x_1,...,x_n) \in {\mathbb B}^{n}$, and ${\bf V}=\{1,2,...,n\}$ be the set of indices of ${\bf x}$. We introduce a \emph{set representation} to denote the labelings of ${\bf x}$. Let $S_4=\{1,2,3,4\}$ and let ${\cal P}$ be the power set of $S_4$. For example, a labeling $\{x_1=1,x_2=0,x_3=1,x_4=1\}$ is denoted by the set $\{1,3,4\}$. For a subset $A \subseteq V$, let us denote by $\mathbf{1}^A \in {\mathbb B}^n$ its characteristic vector, i.e.

\begin{equation}
\mathbf{1}^S_j = \begin{cases} 1 & \text{if~$j \in A,$}\\
0 & \text{otherwise.} 
\end{cases}
\label{eq.charvec}
\end{equation} 

\begin{definition}
Submodular functions map
$f:{\mathbb B}^n \to {\mathbb R}$ and satisfy the following condition:
\begin{equation}
f(X)+f(Y) \geq f(X \vee Y)+f(X \wedge Y),
\label{eq.submodularity2}
\end{equation}
where $X$ and $Y$ are elements of ${\mathbb B}^n$ and the symbols $\vee$ and 
$\wedge$ denote union and intersection of sets respectively. 

\label{def.sbf}
\end{definition}
\Vspace{-0.2cm} In this paper, we use a pseudo-Boolean polynomial
representation for denoting submodular functions.

\begin{definition}
Pseudo-Boolean functions ({\sc pbf}) take a Boolean vector as argument and return a real number, 
i.e. $f: {\mathbb B}^n \to {\mathbb R}$ \cite{BorosH02}. These can be uniquely expressed as multi-linear 
polynomials, i.e. for all $f$ there exists a unique set of real numbers $\{a_S :S \in {\mathbb B}^n\}:$
\begin{equation}
\label{eq:standard}
f(x_1,...,x_n)=\sum_{S \subseteq V}a_S( \prod_{j \in S}x_j), ~~~~~~~~~~a_S \in
{\mathbb R},
\end{equation}
where $a_{\emptyset}$ is said to be the constant term. 
\label{def.pbf}
\end{definition}
The term \emph{order} refers to the maximum
degree of the polynomial. A submodular function of second order
involving Boolean variables can be easily represented using a graph
such that the minimum cut, computed using a max-flow algorithm, also
efficiently minimizes the function. However, max-flow algorithms can
not exactly minimize non-submodular functions or some submodular ones
of an order greater than 3~\cite{zivny09a}.
There is a long history of research in solving subclasses of
submodular functions both exactly and efficiently using max-flow
algorithms~\cite{billionnet85,kolPAMI04,hammer65,zalesky03,queyranne95}. 
In this paper, we propose a linear programming formulation that is capable 
of answering this question: given any pseudo Boolean function, it 
can derive a quadratic submodular formulation of the same cost or a {\em closest} 
quadratic submodular function (i.e., say under $L_1$ norm), if an exact derivation 
does not exist. The problem of using a linear program (LP) for expressing a given 
function using other functions (with \av s) was already established in 
~\cite{cohen06}. Compared to the existing results, we also provide a smaller LP 
for submodular functions and also show that we need only fewer \av s compared to 
existing methods. 



\begin{definition}
  ${\cal F}^k$ denotes a class of pseudo-Boolean functions of order $k$ such that every function $f({\bf x}) \in {\cal F}^k$ 
	satisfies the submodularity property given in Definition~\ref{eq.submodularity2}.
\end{definition}

It was first shown in \cite{hammer65} that any function in ${\cal F}^2$ can be minimized exactly using a max-flow algorithm. Billionnet and Minoux~\cite{billionnet85} showed that any function in ${\cal F}^3$ can be transformed into a function in ${\cal F}^2$ using additional variables. 
While transforming a given higher order function to a function in ${\cal F}^2$, we use additional variables that we refer to as \emph{auxiliary variables} (\av ). In the course of this paper, you will see that these \av s are often more difficult to handle than  variables in the original function and our algorithms are driven by the quest to understand the role of these auxiliary variables and to eliminate the unnecessary ones. 

Kolmogorov~\cite{Kolmogorov2012} improved the complexity of Iwata's capacity scaling algorithm \cite{iwata03} for special functions which are represented 
as a sum of submodular terms. This is the first line of research that does not use auxiliary variables to handle higher order terms. The formulation of 
Kolmogorov also closely resembles the approach of Cooper \cite{cooper2008constraints}, who used a linear program with an exponential number of constraints 
for solving the minimization of the submodular function. It was shown that we can have a algorithm that can be parallelized for minimizing decomposable submodular functions, which can be decomposed into sum of simple submodular functions. In ~\cite{Nishihara2014}, it was shown that the algorithm converges linearly, and they also provide upper and 
lower bounds on the rate of convergence. 

Recently, Zivny et al.~\cite{zivny09a} made substantial progress in characterizing the class of functions that can be transformed to ${\cal F}^2$. Their most notable result is to show that not all functions in ${\cal F}^4$ can be transformed to a function in ${\cal F}^2$. This result stands in strong contrast to the third order case that was positively resolved more than two decades earlier \cite{billionnet85}. Using Theorem 5.2 from \cite{promislow05} it is possible to decompose a given submodular function in ${\cal F}^4$ into 10 different groups ${\cal G}_i,~i=\{1..10\}$, where each ${\cal G}_i$ is shown in Table~\ref{tb.union_of_Gs}. Zivny et al. showed that one of these groups (${\cal G}_{10}$) can not be expressed using any function in ${\cal F}^2$ employing any number of \av s. Most of these results were obtained by mapping the submodular function minimization to a valued constraint satisfaction problem.

\subsection{Problem Statement and main contributions}
\label{sec.problem_statement}
\begin{figure}[!htbp]\centering
\psfig{figure=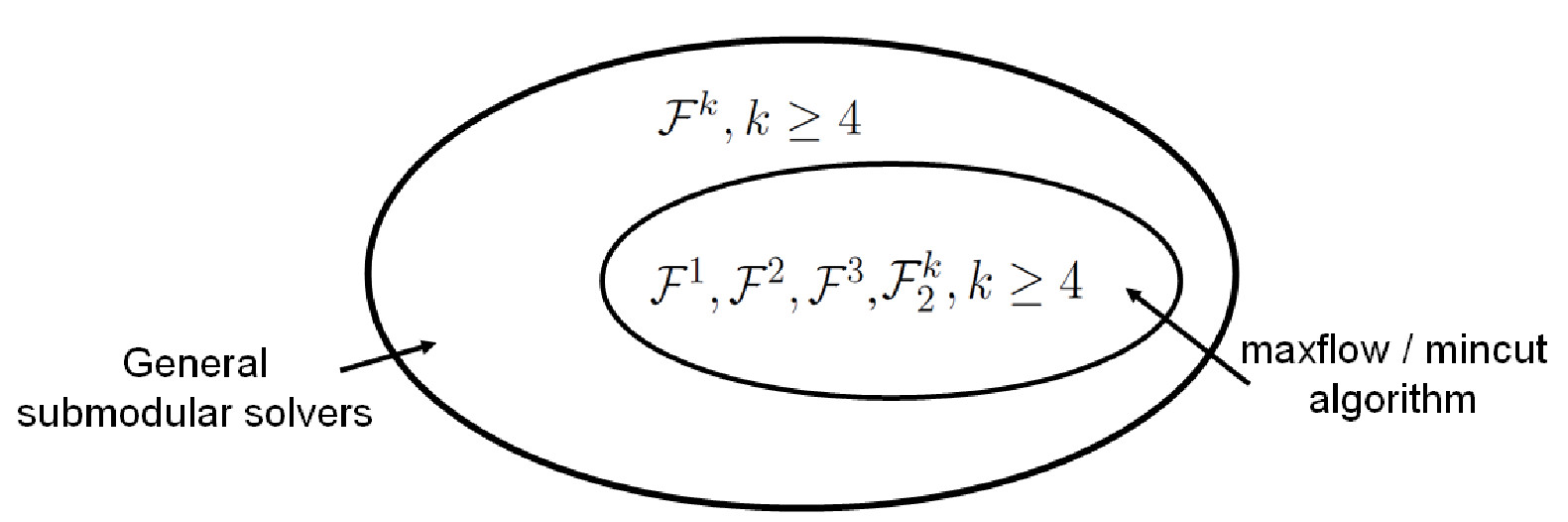,width=0.60\columnwidth}
\caption{\it All the function in the classes ${\cal F}^1,{\cal F}^2,{\cal F}^3$ and 
${\cal F}^k_2,k \ge 2$ can be transformed to functions in ${\cal F}^2$ and minimized 
using the maxflow/mincut algorithm.}
\label{fg.venn_diagram_pbf}
\end{figure}

\paragraph{Largest subclass of submodular functions:}
We are interested in transforming a given function in ${\cal F}^k$
into a function in ${\cal F}^2$ using \av s. As such a transformation
is not possible for all submodular functions of order four or more
\cite{zivny09a}, our goal is to implicitly map the largest subclass
${\cal F}^k_2$ that can be transformed into ${\cal F}^2$.  This
distinction between the two classes ${\cal F}^k_2$ and ${\cal F}^k$
will be crucial in the remainder of the paper (see Figure~\ref{fg.venn_diagram_pbf}).

\begin{definition}
  The class ${\cal F}^k_2$ is the largest subclass of ${\cal F}^k$
  such that every function $f({\bf x}) \in {\cal F}^k_2$ has an
  equivalent quadratic function $h({\bf x},{\bf z}) \in {\cal F}^2$
  using \av s ${\bf z}={z_1,z_2,...,z_m} \in {\mathbb B}^m$
  satisfying the following condition:
\begin{equation}
f({\bf x})=\min_{{\bf z}\in{\mathbb B}^m}h({\bf x},{\bf z}),~~~\forall
{\bf x}.
\label{eq.transformation}
\end{equation}
\end{definition}
In this paper, we are interested in developing an algorithm to
transform every function in this class ${\cal F}^k_2$ to a function in
${\cal F}^2$.

\paragraph{Efficient transformation of higher order functions:}
We propose a linear programming algorithm to transform higher order submodular
functions to quadratic ones using monotonic Boolean 
functions (\mbf ~\cite{Yves2011}). 
This framework provides several advantages. First we show that the
state of an \av\ in a minimum cost labeling is equivalent to an 
\mbf\ defined over the original variables.  This provides an upper bound on
the number of \av s given by the Dedekind number \cite{korshunov81},
which is defined as the total number of \mbf s over a set of $n$
binary variables. In the case of fourth order functions, there are 168
such functions. Using the properties of \mbf s and the nature of these
\av s in our transformation, we prove that these 168 \av s can be
replaced by two \av s.

\paragraph{Minimal use of \av s:}
One of our goals is to use a minimum number ($m$) of \av s in performing the
transformation of~\eqref{eq.transformation}.
Although, given a fixed choice of ${\cal F}^k_2$, reducing the value of $m$ 
does not change the complexity of the resulting min/cut algorithm asymptotically, 
it is crucial in several machine learning and computer vision problems. In general, 
most image based labeling problems involve millions of pixels and in typical problems,
the number of fourth order priors is linearly proportional to the number of pixels. 
Such problems may be infeasible for large values of $m$. It was shown that the 
transformation of functions in ${\cal F}^4_2$ can be achieved using about 30 auxiliary 
variables \cite{zivny09b}. On the other hand, we show that we can transform the same class 
of functions using only 2 additional nodes. Note that this reduction is applicable
to every fourth order term in the function. A typical vision problem may involve 
functions having $10000$ ${\cal F}^4_2$ terms for an image of size $100\times100$. 
Under these parameters, our algorithm will use $20000$ \av s, whereas the existing 
approach \cite{zivny09b} would use as many as 300000 \av s. In several practical 
problems, this improvement will make a significant difference in the running time of
the algorithm.

For a function in ${\cal F}^k_2$, the maximum number of \av s required is given by $2^{2^k}$~\cite{cohen06}. 
We show that one can transform the function using substantially fewer number of \av s given 
by Dedekind number. In section~\ref{sec.auxMBF}, we show that the Dedekind number is substantially 
lower than $2^{2^k}$. In \cite{cohen06}, an LP based approach was used to obtain the bound of $2^{2^k}$. 
We also use an LP-based approach, however the use of monotonic Boolean functions enables us 
to improve this bound to Dedekind number. The idea of reducing the number of \av s in an LP formulation 
has been done in other contexts \cite{trevisan2000SIAM}. In \cite{trevisan2000SIAM}, a combinatorial structure 
commonly referred to as \emph{gadgets} were computed using linear programming. This enables the transformation of 
constraints from one optimization problem to another. In this work, we show that we can transform a function with several 
\av s to a function involving much fewer \av s using a linear programming approach.

\subsection{Limitations of Current Approaches and Open Problems}
\paragraph{Decomposition of submodular functions:}
Many existing algorithms for transforming higher order functions target the minimization of a single $k$-variable \kth order
function. However, the transformation framework is incomplete without showing that a given $n$-variable submodular function of \kth
order can be decomposed into several individual $k$-variable \kth order sub-functions. Billionnet proved that it is possible to decompose
a function in ${\cal F}^3$ involving several variables into 3-variable functions in ${\cal F}^3$ \cite{billionnet85}. To the best of our
knowledge, the decomposition of fourth or higher order functions is still an open problem and it will remain a hard problem due to the following reasoning. In \cite{gallo89}, it was proven that testing a membership of a function $f$ with $n$ variables in $\mathcal{F}^4$ is NP-complete. It is easy to test the submodularity of a fourth order function with $4$ variables. Thus if a function $f$ with $n$ variables is decomposed into several 4-variable fourth order functions and if each of these individual 4-variable functions are submodular, then the function $f$ is submodular. This seems to be most possible case when we know that a function is submodular. Thus it is very unlikely to know that a function is submodular and not know its decomposition. Given this, it is likely that specialized solvers based on max-flow algorithms may never solve the general class of submodular functions. However, this decomposition problem is not a critical issue in machine learning and vision problems. This is because the higher order priors from natural statistics already occur in different sub-functions of $k$ nodes - in other words, the decomposition is known a priori. This paper only focuses on the transformation of a single $k$-variable function in ${\cal F}^k$. As mentioned above, the solution to this problem is still sufficient to solve large functions with hundreds of nodes and higher order priors in applications.

\paragraph{Non-Boolean problems:}
The results in this paper are applicable only to set or pseudo-Boolean functions. Many real world problems involve variables that can take multiple discrete values. Ishikawa showed that it is possible to transform a multi-label second order function to a Boolean second order function using Boolean variables to encode multi-label variables~\cite{bb30984}. To denote a single multi-label variable with $l$ labels, $l$ Boolean variables were used. Ishikawa's method considered functions with convex priors, a class of functions that is slightly more restricted than general submodular functions. Schlesinger and Flach later showed that it is possible to transform general submodular multi-label functions of second order to Boolean second order functions~\cite{schlesinger06}. This approach used $l-1$ Boolean variables to encode an $l$-label multi-label variable. Ramalingam et al.~\cite{Ramalingam08a} generalized this work for transforming multi-label higher order functions to Boolean second-order functions. In~\cite{Ramalingam08a}, the transformation does not preserve submodularity for fourth or higher order functions~\cite{Ramalingam08a}. Zivny et al.~\cite{zivny09a} proved that it is not possible to have a submodularity preserving transformation for fourth or higher order functions.

\paragraph{Excess \av s:}
The complexity of an efficient max-flow algorithm is $O((n+m)^3)$ where $n$ is the number of variables in the original higher order function and $m$ is the number of \av s. Typically in imaging problems, the number of higher order terms is of $O(n)$ and the order $k$ is less than 10. Thus the minimization of the function corresponding to an entire image with $O(n)$ higher order terms will still have a complexity of $O((n+n)^3)$. However when $m$ becomes  at least quadratic in $n$, for example, if a higher-order term is defined over every triplet of variables in $V$, the complexity of the max-flow algorithm will exceed that of a general solver being $O((n+n^3)^3)$. Thus in applications involving a very
large number of higher order terms, a general solver may be more appropriate.

\Vspace{-0.6cm}
\section{Preliminaries}
\label{sec.preliminaries}


\begin{definition}
The (discrete) derivative of a function $f(x_1,\ldots,x_n)$ with respect to $x_i$ is given by:
\begin{equation}
\frac{\delta f}{\delta x_i}(x_1,\ldots,x_n)=f(x_1,\ldots,x_{i-1},1,x_{i+1},\ldots,x_n)-f(x_1,\ldots,x_{i-1},0,x_{i+1},\ldots,x_n).
\end{equation}
\label{def.discrete_derivative}
\end{definition}
\begin{definition}
The second discrete derivative of a function $\Delta_{i,j}({\bf x})$ is given by
\begin{align}
\Delta_{i,j}({\bf x})&=\frac \delta {\delta x_j} \frac{\delta f}{\delta x_i}(x_1,\ldots,x_n)\label{2nd.dev.eq}\\
&\scriptstyle=
\Big(f(x_1,\ldots,x_{i-1},1,x_{i+1}\ldots,x_{j-1},1,x_{j+1}\ldots,x_n)-
f(x_1,\ldots,x_{i-1},0,x_{i+1}\ldots,x_{j-1},1,x_{j+1}\ldots,x_n)\Big)\nonumber\\
&\scriptstyle -\Big(
f(x_1,\ldots,x_{i-1},1,x_{i+1}\ldots,x_{j-1},0,x_{j+1}\ldots,x_n)-
f(x_1,\ldots,x_{i-1},0,x_{i+1}\ldots,x_{j-1},0,x_{j+1}\ldots,x_n)\Big).\nonumber
\end{align}
Note that it follows from the definition of submodular functions \eqref{eq.submodularity2}, that their second derivative is always non-positive for all $\bf x$.
\end{definition}

\section{Transforming functions in ${\cal F}^n_2$ to ${\cal F}^2$}
\label{sec.transformation2}
Consider the following submodular function $f({\bf x}) \in {\cal F}^n_2$ represented as a multi-linear polynomial:
\begin{equation}
f({\bf x})=\sum_{S \in {\mathbb B}^n}a_S( \prod_{j \in S}x_j), ~~~~~a_S \in
{\mathbb R}.
\label{eq.function_gg}
\end{equation}


Let us consider a function $h({\bf x},{\bf z}) \in {\cal F}^2$ where $\bf z$ is a set of \av s used 
to model functions in ${\cal F}^n_2$. Any general function in ${\cal F}^2$ can be represented as a 
multi-linear polynomial (consisting of linear and bi-linear terms involving all variables):
\begin{equation}
h({\bf x},{\bf z})=\sum_{i} a_i\,x_i -\sum_{i,j:i>j} a_{i,j}\,x_i x_j
+\sum_{l} a_l\,z_l -\sum_{l,m:l>m} a_{l,m}\,z_l z_m  -\sum_{i,l} a_{i,l}\,x_iz_l.
\label{eq.function_hhh}
\end{equation} 

The negative signs in front of the bi-linear terms $(x_i x_j,z_l x_i,z_l z_m)$ emphasize that their 
coefficients ($-a_{ij},-a_{i l},-a_{l m}$) must be non-positive if the function is
submodular. We are seeking a function $h$ such that:
\begin{equation}
f({\bf x})=\min_{{\bf z}\in{\mathbb
    B^n}}h({\bf x},{\bf z}),\forall {\bf x}.
\label{eq.function_fff}
\end{equation}

\noindent
Here the function $f({\bf x})$ is known. We are interested in
computing the coefficients ($\bf a$), and in determining the number of
auxiliary variables required to express a function as a pairwise
submodular function. The problem is challenging due to the
inherent instability and dependencies within the problem -- different
choices of parameters cause auxiliary variables to take different
states. To explore the space of possible solutions fully, we
must characterize what states an \av\ takes.
\subsection{Auxiliary Variables as Monotonic Boolean Functions}
\label{sec.auxMBF}
\begin{definition}
A monotonic (increasing) Boolean function (\mbf) $m:{\mathbb B}^n \to {\mathbb B}$ takes a Boolean vector as argument and 
returns a Boolean, s.t if $y_i \leq x_i,~~\forall i \implies m({\bf y}) \leq m({\bf x})$.
\label{def.mbf}
\end{definition}

\begin{lemma}
Let $z_s ({\bf x})$ be a function that takes an argument ${\bf x}$ and returns a Boolean as shown below:
\begin{equation}
  z_s({\bf x})=\arg \min_{z_s} \left( \min_{\bf z'} h({\bf x},{\bf z}', z_s) \right),
  \label{eq.z_s}
\end{equation}
where $h({\bf x},{\bf z}',z_s)$ is a submodular function defined in Equation~(\ref{eq.function_hhh}) and 
satisfying Equation~(\ref{eq.function_fff}). The function 
$z_s ({\bf x})$ that maps a Boolean vector $\bf x$ to the Boolean state of $z_s$ is an \mbf~(See
Definition~\ref{def.mbf}), where $\bf z'$ is the set of all auxiliary variables except $z_s$.
\label{lem.mbf}
\end{lemma}
\begin{proof}
  We consider a current labeling $\bf x$ with an induced labeling of
  $z_s=z_s({\bf x})$.
We first note 
\begin{equation}
 h'({\bf x},z_s)=\min_{{\bf z}'}  h({\bf x},{\bf z}',z_s)
\end{equation}
is a submodular function i.e. it satisfies
\eqref{eq.submodularity2}. We now consider {\em increasing} the value
of $\bf x$, that is given a current labeling $\bf x$ we consider a new labeling ${\bf x}^{(i)}$ such that
\begin{equation}
x^{(i)}_j=\begin{cases} 1&\text{ if $j=i$ }\\
x_j &\text{ otherwise.}
\end{cases}
\end{equation}
We wish to prove
\begin{equation}
z_s({\bf x}^{(i)}) \geq z_s({\bf x})\,\, \forall {\bf x},i .
\end{equation}
Note that if $z_s({\bf x})=0$ or $x_i=1$ this result is trivial. This leaves 
the case:  $z_s({\bf x})=1$ and $x_i=0$. It follows from \eqref{2nd.dev.eq} that:
\begin{align}
h'(x_1,\ldots,x_{i-1},1,x_{i+1},\ldots,0)&-
h'(x_1,\ldots,x_{i-1},0,x_{i+1},\ldots,0) \geq \nonumber \\
h'(x_1,\ldots,x_{i-1},1,x_{i+1},\ldots,1)&-
h'(x_1,\ldots,x_{i-1},0,x_{i+1},\ldots,1).
\label{eqn_subm_2}
\end{align}
Using Equation~(\ref{eq.z_s}), we derive the following from our hypothesis $z_s({\bf x})=1$ and $x_i=0$:
\begin{equation}
h'(x_1,\ldots,x_{i-1},0,x_{i+1},\ldots,0) \geq h'(x_1,\ldots,x_{i-1},0,x_{i+1},\ldots,1).
\end{equation}
Hence by replacing $h'(x_1,\ldots,x_{i-1},0,x_{i+1},\ldots,0)$ with $h'(x_1,\ldots,x_{i-1},0,x_{i+1},\ldots,1)$ in 
Equation~\eqref{eqn_subm_2}, we have
\begin{align}
h'(x_1,\ldots,x_{i-1},1,x_{i+1},\ldots,0)&-
h'(x_1,\ldots,x_{i-1},0,x_{i+1},\ldots,0) \geq\\
h'(x_1,\ldots,x_{i-1},1,x_{i+1},\ldots,1)&-
h'(x_1,\ldots,x_{i-1},0,x_{i+1},\ldots,0).\nonumber
\end{align}
This implies the following:
\begin{equation}
h'(x_1,\ldots,x_{i-1},1,x_{i+1},\ldots,0)\geq h'(x_1,\ldots,x_{i-1},1,x_{i+1},\ldots,1) .
\end{equation}
Therefore $z_s({\bf x}^{(i)})=1$ as per the Equation~(\ref{eq.z_s}).  Repeated application of the
statement gives $y_i \leq x_i, \forall i \implies z_s({\bf y}) \leq
z_s({\bf x})$ as required\qed
\end{proof}

\begin{definition}
  The Dedekind number $M(n)$ is the number of \mbf s of n
  variables. Finding a closed-form expression for $M(n)$ is known as
  the Dedekind problem \cite{kleitman69,korshunov81}.
\label{def.dedekind}
\end{definition}
The Dedekind number of known values are shown below: $M(1)=3$, this
corresponds to the set of functions:
\begin{equation}
  M_1(x_1)\in \{ {\bf 0},{\bf 1}, x_1 \},
\end{equation}
where $\bf 0$ and $\bf 1$ are the functions that take any input and
return 0 or 1 respectively.  $M(2)=6$ corresponding to the set of
functions:
\begin{equation}
M_2(x_1,x_2)=\{{\bf 0},{\bf 1},x_1,x_2,x_1 \vee x_2, x_1 \wedge x_2\}.
\end{equation}

Similarly, $M(3)=20$, $M(4)=168$, $M(5)=7581$, $M(6)\approx 7.8 \times 10^6$,
$M(7)\approx 2.4\times 10 ^{12}$, and $M(8)\approx 5.6\times 10^{23}$. 

\begin{theorem}
On transforming the largest graph-representable subclass of \kth order function to pairwise Boolean function, the upper 
bound on the maximal number of required \av s is given by the Dedekind number $D(k)$.
\label{th.bounds}
\end{theorem}
\begin{proof}
  The proof is straightforward. Consider a general multinomial, of
  similar form to Equation \eqref{eq.function_gg} with more than
  $D(k)$ \av s. It follows from Lemma \ref{lem.mbf} that at least
  2 of the \av s must correspond to the same \mbf, and always
  take the same values.  Hence, all references to one of these {\sc
    av} in the pseudo-Boolean representation can be replaced with
  references to the other, without changing the associated costs.
  Repeated application of this process will leave us with a solution
  with at most $D(k)$ \av s.\qed
\end{proof}
Although this upper bound is large for even small values of
$k$, it is much tighter than the existing upper bound of $S(k)=2^{2^k}$\cite{cohen06} (also see 
Proposition 24 in \cite{zivny08Report}). 

\begin{lemma}
Let $D(k)$ denote the Dedekind number for all positive values of $k$. Given $S(k)=2^{2^k}$ and for 
even values of $k$, we have:
\begin{equation}
S(k) \ge 2^{\sum_{i \in \{0,1,...,k\} \backslash \{\frac{k}2 - 1,\frac{k}2\}}{k \choose i}}D(k).
\end{equation}
When $k$ is odd, we have:
\begin{equation}
S(k) \ge 2^{\sum_{i \in \{0,1,...,k\} \backslash \{\frac{k-1}2,\frac{k+1}2\}}{k \choose i}}D(k).
\end{equation}

\label{lem:ded-upperbound}
\end{lemma}
\begin{proof}
For even small values of $k=\{3,...,8\}$ the upper bound using Dedekind's number is much tighter compared 
to $S(k)$:$(M(3)=20,S(3)=256),~(M(4)=168,S(4)=65536),~(M(5)=7581,S(5)\approx 4.29\times10^9),~(M(6)\approx 7.8 \times 10^6,S(6) \approx 1.85 \times 10^{19}),~(M(7)\approx 2.4\times 
10 ^{12},S(7) \approx 3.4 \times 10^{38})$, and $(M(8)\approx 5.6\times 10^{23},S(8)\approx 1.156 \times 10^{77})$. For $k>8$, $D(k)$ remains unknown, and the development of a closed form solution remains an active area of research.

Several upper bounds have been derived for $D(k)$ and we use the following bound by 
Hansel \cite{hansel66,kleitman69} to prove our result.
\begin{equation}
D(k) \le 3^{k \choose {\lfloor \frac{k}{2} \rfloor}},
\label{eq.hanselbound}
\end{equation}

\begin{equation}
D(k) \le 2^{\log_2(3){k \choose {\lfloor \frac{k}{2} \rfloor}}}.
\label{eq.hanselbound}
\end{equation}
The proof is given for two different 
cases depending on whether $k$ is even or odd. First let us consider the case when $k$ is even.
\begin{equation}
{k \choose {\lfloor \frac{k}{2} \rfloor}}={k \choose \frac{k}{2}}.
\end{equation}
We can obtain the following:
\begin{equation}
{k \choose \frac{k}{2}}=\frac{k\times(k-1)...(k-\frac{k}{2})}{1 \times 2 ... (\frac{k}{2})}=
\frac{k\times(k-1)...(k-(\frac{k}{2}-1))}{1 \times 2 ... (\frac{k}{2}-1)}={k \choose (\frac{k}{2}-1)}.
\label{eq.even_kminus1}
\end{equation}
Using binomial theorem we know that 
\begin{equation}
\sum_{i \in \{0,1,...k\}}{k \choose i}=2^k.
\label{eq.binth}
\end{equation}
Using Equations(~\ref{eq.even_kminus1}) and (~\ref{eq.binth}) we have the following Equation:
\begin{equation}
2^k=\sum_{i \in \{0,1,...,\frac{k}{2}-2,\frac{k}{2}+1,...,k\}}{k \choose i}+2{k \choose \frac{k}{2}}.
\end{equation}
Since $\log_2(3) < 2$, it is easy to observe the following:
\begin{equation}
2^k \ge \sum_{i \in \{0,1,...,\frac{k}{2}-2,\frac{k}{2}+1,...,k\}}{k \choose i}+\log_2(3){k \choose \frac{k}{2}}.
\end{equation}
Taking both sides to the exponent of 2, we have the following:
\begin{equation}
2^{2^k} \ge 2^{\sum_{i \in \{0,1,...,\frac{k}{2}-2,\frac{k}{2}+1,...,k\}}{k \choose i}+\log_2(3){k \choose \frac{k}{2}}}
\end{equation}
\begin{equation}
2^{2^k} \ge 2^{\sum_{i \in \{0,1,...,\frac{k}{2}-2,\frac{k}{2}+1,...,k\}}{k \choose i}}2^{\log_2(3){k \choose \frac{k}{2}}}
\end{equation}
\begin{equation}
S(k) \ge 2^{
\sum_{i \in \{0,1,...,k\} \backslash \{\frac{k}2 - 1,\frac{k}2\}}
{k \choose i}}D(k).
\end{equation}
This implies that $S(k)$ is significantly larger than $D(k)$. Let us consider the case when $k$ is odd. 
\begin{equation}
{k \choose {\lfloor \frac{k}{2} \rfloor}}={k \choose \frac{k-1}{2}}.
\end{equation}
It is well known that:
\begin{equation}
{k \choose \frac{k+1}{2}}=\frac{k\times(k-1)...(k-\frac{k-1}{2})(k-\frac{k+1}{2})}{1 \times 2 ... (\frac{k-1}{2})(\frac{k+1}{2})}=
{k \choose \frac{k-1}{2}}\frac{k-1}{k+1}={k \choose \frac{k-1}{2}}(1-\frac{2}{k+1}).
\label{eq.odd_kplus1}
\end{equation}
Using Equations(~\ref{eq.odd_kplus1}) and (~\ref{eq.binth}) we have the following Equation:
\begin{equation}
2^k=\sum_{i \in \{0,1,...,\frac{k-3}{2},\frac{k+3}{2},...,k\}}{k \choose i}+(1+1-\frac{2}{k+1}){k \choose \frac{k-1}{2}}.
\end{equation}
Since $\log_2(3) < (1+1-\frac{2}{k+1})$ for $k>8$, it is easy to observe the following:
\begin{equation}
2^k \ge \sum_{i \in \{0,1,...,\frac{k-3}{2},\frac{k+3}{2},...,k\}}{k \choose i}+\log_2(3){k \choose \frac{k-1}{2}}.
\end{equation}
By lifting both sides to the power of 2, we have the following relation:
\begin{equation}
2^{2^k} \ge 2^{\sum_{i \in \{0,1,...,\frac{k-3}{2},\frac{k+3}{2},...,k\}}{k \choose i}+\log_2(3){k \choose \frac{k-1}{2}}}
\end{equation}
\begin{equation}
2^{2^k} \ge 2^{\sum_{i \in \{0,1,...,\frac{k-3}{2},\frac{k+3}{2},...,k\}}{k \choose i}}2^{\log_2(3){k \choose \frac{k-1}{2}}}
\end{equation}
\begin{equation}
S(k) \ge 2^{
\sum_{i \in \{0,1,...,k\} \backslash \{\frac{k-1}2,\frac{k+1}2\}}
{k \choose i}}D(k).
\end{equation}
\qed
We observe that $S(k)$ is significantly larger than $D(k)$ when $k$ is odd.
\end{proof}

In \cite{zivny10constraints}, the problem of improving this upper bound was mentioned as an open 
problem. 
In some sense, both these upper bounds are not practically feasible for even small values of 
$k$. This number is prohibitive because we are looking for an exact transformation that preserves submodularity. 
By using auxiliary variables, we can also transform a given higher order function to a non-submodular one using 
much fewer variables \cite{ishikawa2011,fix11,gallagher11}. In section \ref{sec:tight-bounds}, we will further tighten the 
bound for fourth order functions.

Note that this representation of \av s as \mbf~is over-complete, for
example if the \mbf~of a auxiliary variable $z_i$ is the constant
function $z_i({\bf x})=\bf 1$ we can replace $\min_{{\bf z},z_i}h({\bf
  x},{\bf z}, z_i)$ with the simpler (i.e. one containing less
auxiliary variables) function $\min_{{\bf z}}h({\bf x},{\bf z},
1)$. 

Given any function $f$ in ${\cal F}^k_2$, the equivalent pairwise 
form $f'\in {\cal F}^2$ can be found by solving a linear program.
The construction of the linear program is given in the following section.

\section{The Linear Program}
\label{sec.general_case}
A sketch of the formulation can be given as follows: In general, the
presence of \av s of indeterminate state, given a labeling {\bf x}
makes the minimizing an LP non-convex and challenging to solve
directly. Instead of optimizing this problem containing \av s of
unspecified state, we create an auxiliary variable associated
with every \mbf. Hence given any labeling $\bf x$ the state of
every auxiliary variable is fixed a priori, making the problem
convex. We show how the constraints that a particular \av\ must
conform to a given \mbf~can be formulated as linear constraints,
and that consequently the problem of finding the closest member of
$f'\in {\cal F}^2 $ to any pseudo Boolean function is a linear
program.

This program will make use of the max-flow linear program formulation
to guarantee that the minimum cost labeling of the \av s corresponds 
to their \mbf s. To do this we must first rewrite the cost of 
Equation~(\ref{eq.function_hhh}) in a slightly different form. We write:
\begin{align}
 f({\bf x},{\bf z})&= c_\emptyset +\sum_{i} c_{i,s}\,(1-x_i) +\sum_{i} c_{t,i}\,x_i   +\sum_{i,j:i>j} c_{i,j}\,x_i (1-x_j)\nonumber\\
 &+\sum_{l} c_{l,s} \,(1-z_l)+\sum_{l} c_{t,l} \,z_l +\sum_{l,m:l>m} c_{l,m}\,z_l\,(1-z_m)  +\sum_{i,l} c_{i,l}\,x_i\,(1-z_l).
\label{eq.gc_cost}
\end{align}
where $c_\emptyset$ is a constant that may be either positive or negative and all other $c$ are non-negative values referred to as the {\em capacity} of an edge. By~\cite{fisher1978}, this form is equivalent to that of (\ref{eq.function_hhh}), in that any function that can be written in form (\ref{eq.function_hhh}), can also be written as (\ref{eq.gc_cost}) and visa versa.
\subsection{The Max-flow Linear Program}
Under the assumption that $\bf x$ is fixed, we are interested in
finding a minimum of the Equation:
\begin{align}
 f_{\bf x}({\bf z})&=~c_\emptyset +\sum_{i} c_{i,s}\,(1-x_i) +\sum_{i} c_{t,i}\,x_i   +\sum_{i,j:i>j} c_{i,j}\,x_i (1-x_j)\nonumber\\
 &~+\sum_{l} c_{l,s} \,(1-z_l)+\sum_{l} c_{t,l} \,z_l
 +\sum_{l,m:l>m} c_{l,m}\,z_l\,(1-z_m)  +\sum_{i,l}
 c_{i,l}\,x_i\,(1-z_l)\nonumber\\
&=~d_{{\bf x},\emptyset} +\sum_{l} d_{{\bf x},l,s} \,(1-z_l)+\sum_{l} d_{{\bf x},t,l} \,z_l  +\sum_{l,m:l>m} d_{{\bf x},l,m}\,z_l\,(1-z_m) 
\label{eq.flow}
\end{align}
where
\begin{equation}
 d_{{\bf x},\emptyset} = c_\emptyset +\sum_{i: x_i=0} c_{i,s} +\sum_{i:x_i=1}
 c_{t,i}   +\sum_{i,j:i>j \wedge x_i=1 \wedge x_j=0} c_{i,j}
\end{equation}
\begin{equation}
 d_{{\bf x},l,s}=c_{l,s} +\sum_{i : x_i=1} c_{i,l},
 \text{ } d_{{\bf x},t,l}=c_{t,l} \text{ and } d_{{\bf x},l,m}=c_{l,m}.
\end{equation}
Then the minimum cost of Equation~(\ref{eq.gc_cost}) may be found by
solving its dual max-flow program. Writing $\nabla_{{\bf x},s}$ for flow from the sink, and $\nabla_{{\bf x},t}$ for flow to the sink, we seek
\begin{equation}
  \max \nabla_{{\bf x},s} +d_{{\bf x},\emptyset},
\end{equation}
subject to the constraints that
\begin{equation}
\begin{array}{rclr} f_{{\bf x},ij} -  d_{{\bf x},ij} & \leq & 0, &
  ~~~~~\forall (i, j) \in E \\
\sum_{j: (j, i) \in E} f_{{\bf x},ji} - \sum_{j: (i, j) \in E} f_{{\bf x},ij} & \leq &
0, & ~~~~~\forall i \neq s,t \\
\nabla_{{\bf x},s} + \sum_{j: (j, s) \in E} f_{{\bf x},js} - \sum_{j: (s, j) \in E} f_{{\bf x},sj} & \leq & 0 & \\
\nabla_{{\bf x},t} + \sum_{j: (j, t) \in E} f_{{\bf x},jt} - \sum_{j: (t, j) \in E} f_{{\bf x},tj} & \leq & 0 & \\
f_{{\bf x},ij} & \geq & 0, & (i, j) \in E\\
\end{array}
\label{eq:cond_flow}
\end{equation}
where $E$ is the set of all ordered pairs $(l,m) : \forall l>m $,
$(s,l): \forall l$ and $(l,t): \forall t$, and $f_{{\bf x},i,j}$
corresponds to the flow through the edge $(i,j)$.

We will not use this exact {\sc lp} formulation, but instead rely on the fact that {\em \mbox{$f_{\bf x}({\bf z})$} is a minimal cost
labeling if and only if there exists a flow satisfying constraints
(\ref{eq:cond_flow}) such that}
\begin{equation}
\label{reduce_dep}
  f_{{\bf x}} ({\bf z}) - \nabla_{{\bf x},s}  -d_{{\bf x},\emptyset} \leq 0.
\end{equation}

\subsection{Choice of \mbf~as a set of linear constraints}
We are seeking minima of a quadratic pseudo Boolean function of the
form \eqref{eq.gc_cost},
where $\bf x$ is the variables we are interested in minimizing and
$\bf z$ the auxiliary variables. As previously mentioned, formulations
that allow the state of the auxiliary variable to vary tend to result in
non-convex optimization problems. To avoid such difficulties, we 
specify as the location of minima of $\bf z$ as a set of hard constraints.
We want that:
\begin{equation}
\label{eq:ineq}
\min_{\bf z} f_{\bf x}({\bf z}) = f_{\bf x}( [m_1 ({\bf x}), m_2 ({\bf x}),\ldots m_{D(k)} ({\bf x})])\,\, ~~~~~\forall {\bf x}.
\end{equation}
where $f_{\bf x}$ is defined as in \eqref{eq.flow}, and  $m_1, \ldots m_{D(k)}$ are the set of all possible \mbf s
defined over $\bf x$.  
By setting all of the capacities $d_{i,j}$ to $0$, it can be seen that
a solution  satisfying \eqref{eq:ineq} must exist. It follows from the
reduction described in Lemma \ref{lem.mbf}, and that all functions
that can be expressed in a pairwise form can also be expressed in a
form that satisfies these restrictions.



We enforce condition \eqref{eq:ineq} by the set of linear constraints
\eqref {eq:cond_flow} and \eqref{reduce_dep} for all possible choices 
of $\bf x$. Formally we enforce the condition
\begin{equation}
\label{reduce_dep2}
  f_{{\bf x}} ([m_1({\bf x}),\ldots,m_{D(k)}({\bf x})]) - \nabla_{{\bf x},s}  -d_{{\bf x},\emptyset} \leq 0.
\end{equation}
Substituting in (\ref{eq.flow}) we have $2^k$ sets of conditions, namely,
\begin{equation}
  \sum_{l} d_{{\bf x},l,s} \,(1-m_l({\bf x}))+\sum_{l} d_{{\bf x},t,l} \,(1-m_l({\bf x}))  +\sum_{l,m:l>m} d_{{\bf x},l,m}\,m_l({\bf x})\,(1-m_m({\bf x}))  - \nabla_{{\bf x},s}  \leq 0,
\end{equation}
subject to the set of constraints \eqref{eq:cond_flow} for all $\bf
x$.  Note that we make use of the max-flow formulation, and not the
more obvious min-cut formulation, as this remains a linear program even
if we allow the capacity of edges $\bf d$\footnote{ In itself $\bf d$
  is just a notational convenience, being a sum of coefficients in
  $\bf c$.}  to vary.

\paragraph{Submodularity Constraints}
We further require that the quadratic function is submodular or
equivalently, the capacity of all edges $c_{i,j}$ be non-negative. This can
be enforced by the set of linear constraints that 
\begin{equation}
c_{i,j}\geq 0,~~~~~~\forall i,j .
\label{LP2}
\end{equation} 
\subsection{Finding the nearest submodular Quadratic Function}
We now assume that we have been given an arbitrary function  $g ({\bf
  x})$ to minimize, that may or may not lie in ${\cal F}^k$. We are
interested in finding the closest possible function in ${\cal F}^2 $
to it. 
To find the closest function to it (under the $L_1$ norm), we minimize:

\begin{align}
\min_{\bf c} \sum_{{\bf x}\in {\mathbb B}^k} \Big| g({\bf x})-&
\min_{\bf z} f({\bf x},{\bf z})\Big|=\\
\min_{\bf c} \sum_{{\bf x}\in {\mathbb B}^k} \Big| g({\bf x})-&
 f({\bf x},{\bf m}({\bf x}))\Big|=\\
\min_{\bf c} \sum_{{\bf x}\in {\mathbb B}^k} \Big| g({\bf x})-&
\big ( c_\emptyset +\sum_{i} c_{i,s}\,(1-x_i) +\sum_{i} c_{t,i}\,x_i   +\sum_{i,j:i>j} c_{i,j}\,x_i (1-x_j)\\
 &+\sum_{l} c_{l,s} \,(1-m_l ({\bf x}))+\sum_{l} c_{t,l} \,(1-m_l
 ({\bf x})) +\sum_{l,m:l>m} c_{l,m}\,m_l ({\bf x})\,(1-m_m ({\bf x}))
 \nonumber\\
& +\sum_{i,l} c_{i,l}\,x_i\,(1-m_l ({\bf x}))  \big)\Big| \nonumber
\label{LPcost}
\end{align}
where ${\bf m}({\bf x})=[m_1({\bf x}),\ldots,m_{D(k)}({\bf x})] $ is the vector of all \mbf s over $\bf x$, and 
subject to the family of constraints set out in the previous
subsection.  Note that expressions of the form $\sum_i | g_i |$ can be
written as $\sum_i h_i$ subject to the linear constraints $h_i>g_i$
and $h_i >-g_i$ and this is a linear program. \qed
\subsection{Discussion}
Several results follow from the linear program described in the previous section.
In particular, if we consider a function $g$ of the same form as
Equation~\eqref{eq:standard} such that
\begin{equation}
\min_{\bf c} \sum_{{\bf x}\in {\mathbb B}^k} \Big| g({\bf x})-
\min_{\bf z} f({\bf x},{\bf z})\Big|=0.
\end{equation}
exactly defines a linear polytope for any choice of $|{\bf x}|=k$, and this result holds for any choice of basis functions.

Of equal note, the convex-concave procedure~\cite{Yuille02theconcave} 
is a generic move-making algorithm that finds local optima by
successively minimizing a sequence of convex (i.e. tractable)
upper-bound functions that are tight at the current location (${\bf
  x}'$). \cite{NarasimhanB05} showed how this could be similarly done
for quadratic Boolean functions, by decomposing them into submodular
and supermodular components. The work~\cite{cooc} showed that any
function could be decomposed into a quadratic submodular function, and
an additional overestimated term. Nevertheless, this decomposition was
not optimal, and they did not suggest how to find a optimal
overestimation. The optimal overestimation which lies in ${\cal F}^2$
for a cost function defined over a clique $\bf g$ may be found by
solving the above {\sc lp} subject to the additional requirements:
%
\begin{align}
   g({\bf x}) & \leq min_{\bf z} f({\bf x},{\bf z}), ~~~~~\forall {\bf x} \neq {\bf x}'\\
   g({\bf x}')& \geq min_{\bf z} f({\bf x}',{\bf z}).
\end{align}
\paragraph{Efficiency concerns:} As we consider larger cliques, it
becomes less computationally feasible to use the techniques discussed
in this section, at least without pruning the number of auxiliary
variables considered. As previously mentioned, constant \av s and \av s that
corresponds to that of a single variable in $x$ i.e. $z_l=x_i$ can be
safely discarded without loss of generality. 
In the following section, we show that a function in ${\cal F}^4_2$ can be 
represented by only two \av s, rather than 168 as suggested by the Dedekind 
number. However, in the general case a minimal form representation eludes us. 
As a matter of pragmatism, it may be useful to attempt to solve the {\sc lp} 
of the previous section without making use of any \av, and to successively 
introduce new variables, until a minimum cost solution is found.

\section{Tighter Bounds: Transforming functions in ${\cal F}^4_2$ to ${\cal F}^2$}
\label{sec:tight-bounds}
\label{sec.transformation}
Consider the following submodular function $f(x_1,x_2,x_3,x_4) \in {\cal F}^4$ represented as a multi-linear polynomial:
\begin{equation}
f(x_1,x_2,x_3,x_4)=a_0+\sum_{i}a_ix_i+
\sum_{i>j}a_{ij}x_ix_j+
\sum_{i>j>k}a_{ijk}x_ix_jx_k+
a_{1234}x_1x_2x_3x_4,~~~~\Delta_{ij}({\bf x})\leq 0 
\label{eq.function_g}
\end{equation}
where $i,j,k \in S_4$ and $\Delta_{ij}({\bf x})$ is the discrete second derivative of $f({\bf x})$ with respect to $x_i$ and $x_j$. 

Consider a function $h(x_1,x_2,x_3,x_4,z_s) \in {\cal F}^2$ where $z_s$ is an \av\ used to model functions in 
${\cal F}_2^4$. In general, we need several \av s to transform a function in ${\cal F}_2^4$ to a function in 
${\cal F}_2$. Any general function in ${\cal F}^2$ using one \av, can be represented as a multi-linear 
polynomial (consisting of linear and bilinear terms involving all five variables):

\begin{equation}
h(x_1,x_2,x_3,x_4,z_s) = b_0 + 
												 \sum_{i}b_ix_i - 
												 \sum_{i>j}b_{ij}x_ix_j + 
												 (g_s-\sum_{i=1}^4g_{s,i}x_i)z_s,~~~~~b_{ij}\geq 0,g_{s,i}\geq 0,i,j\in S_4.
\label{eq.function_h}
\end{equation}

\noindent
The negative signs in front of the bilinear terms $(x_ix_j,z_sx_i)$ emphasize  that their coefficients 
($-b_{ij},-g_{s,i}$) must be non-positive to ensure submodularity. We have the following condition from 
Equation~(\ref{eq.transformation}), given in page~\pageref{eq.transformation}:

\begin{equation}
f(x_1,x_2,x_3,x_4)=\min_{z_s\in{\mathbb
    B}}h(x_1,x_2,x_3,x_4,z_s),~~~~~\forall {\bf x}.
\label{eq.single_z}
\end{equation}

\noindent
Here the coefficients $(a_i,a_{ij},a_{ijk},a_{ijkl})$ in the function $f({\bf x})$ are known. We wish to compute 
the coefficients $(b_i,b_{ij},g_s,g_{s,n})$ where $i,j \in {\bf V},i \neq j,n\in S_4$. If we were given 
$(g_s,g_{s,i})$ then from Equations~(\ref{eq.function_h}) and (\ref{eq.single_z}) we would have: 
\begin{equation}
z_s=
\begin{cases}
    1 & \mbox{if $g_s-\sum_{i=1}^4g_{s,i}x_i < 0$,}\\
    0 & \mbox{otherwise.}
  \end{cases}
\end{equation}

\noindent
Our main result is to prove that any function $h \in {\cal F}^4_2$ can be transformed to a 
function $h'(x_1,x_2,x_3,x_4,z_{t},z_{r}) \in {\cal F}^2$ involving only two auxiliary 
variables $z_{t}$ and $z_{r}$ as stated in Theorem~\ref{th.reduction_to_joint}. 

\begin{figure*}[!htbp]
\begin{center}
\mbox{
\subfigure[]{\psfig{figure=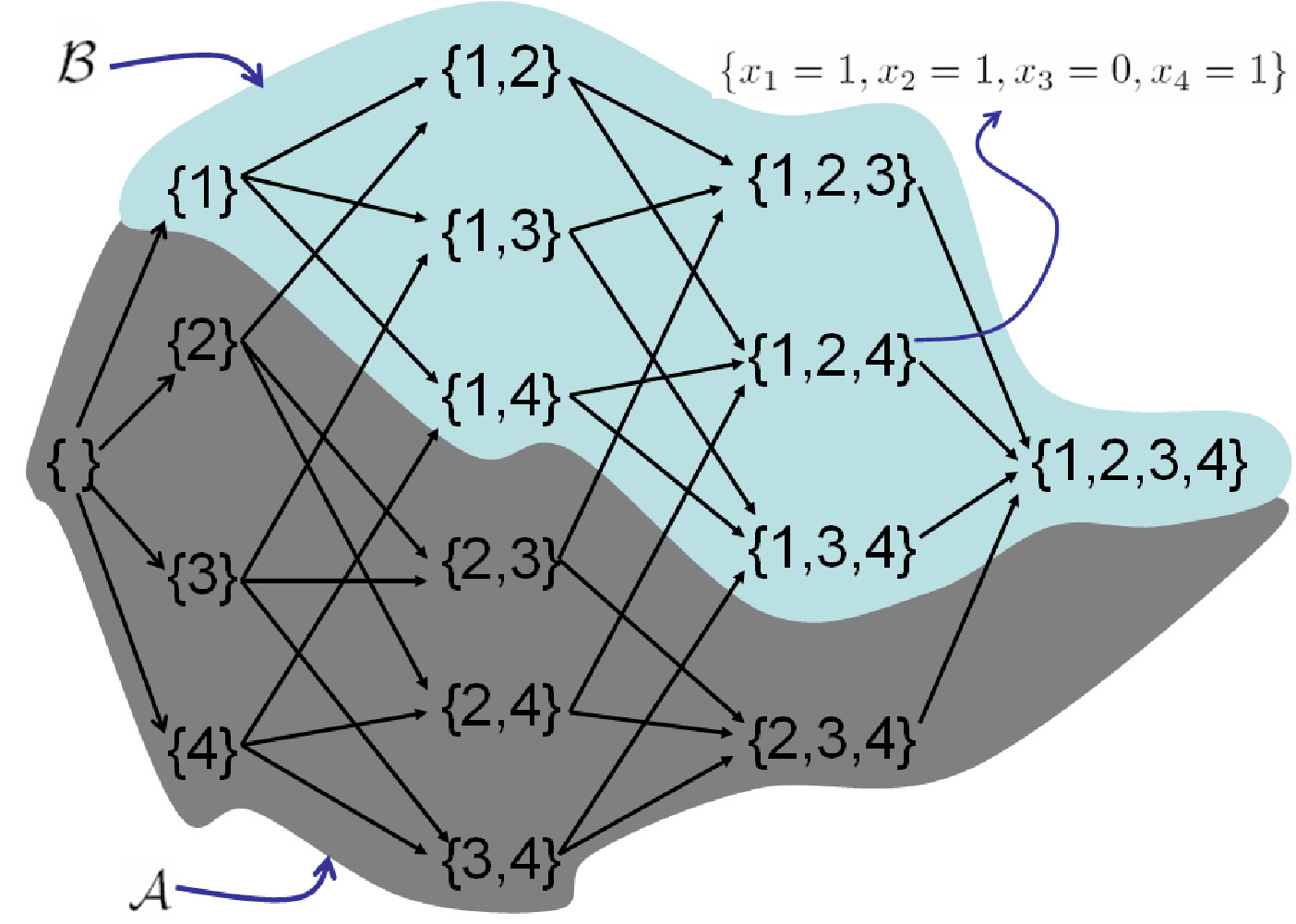,width=0.33\columnwidth}}
\subfigure[]{\psfig{figure=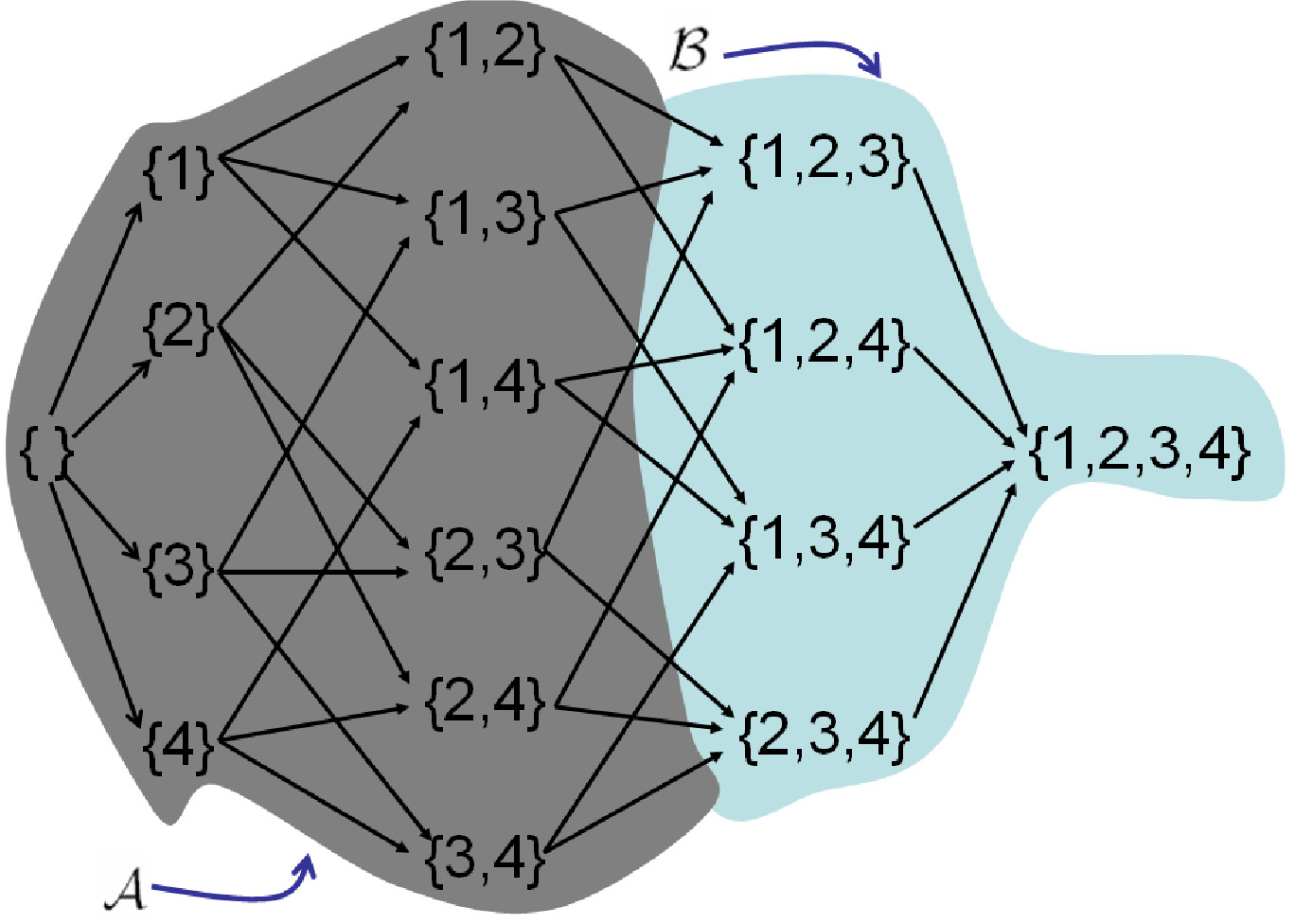,width=0.33\columnwidth}}
}
\end{center}

\caption{We show some examples of partitions using Hasse diagrams. Here, we use set representation 
for denoting the labelings of $(x_1,x_2,x_3,x_4)$. For example the set $\{1,2,4\}$ is equivalent 
to the labeling $\{x_1=1,x_2=1,x_3=0,x_4=1\}$. In (a), ${\cal A}=\{\{\}, \{2\}, \{3\}, \{4\}, 
\{2,3\}, \{2,4\}, \{3,4\}, \{2,3,4\}\}$ and ${\cal B}=\{\{1\}, \{1,2\}, \{1,3\}, \{1,4\}, 
\{1,2,3\}, \{1,2,4\}, \{1,3,4\}, S_4\}$. (a) and (b) are examples of partitions. Any arbitrary 
\av\ must be associated with one of these 168 partitions as given by the Dedekind number $D(k)$. 
}
\label{fg.set_partition}
\end{figure*}
Let ${\cal A}$ be the family of sets corresponding to labelings of
${\bf x}$ such that: $z_s=0=\arg\min_{z_s} h({\bf x},z_s)$. In the same
way let ${\cal B}$ be the family of sets corresponding to labelings
of ${\bf x}$ such that:$z_s=1=\arg \min_{z_s}h({\bf x},z_s)$.
These sets $\cal A$ and ${\cal B}$ partition $\bf x$, as defined below:


\begin{definition}
A partition divides ${\cal P}$ into sets ${\cal A}$ and ${\cal B}$ such that 
${\cal A}=\{{\cal S}({\bf x}):0=\arg\min_{z\in{\mathbb B}} h({\bf x},z),
{\bf x} \in {\mathbb B}^4\}$ and ${\cal B}={\cal P}\backslash{\cal A}$. 
Note that $\emptyset \in {\cal A}$. Here ${\cal S}({\bf x})$ denotes the 
set corresponding to ${\bf x}$.
\label{def.partition}
\end{definition}

\noindent
In the rest of the paper, we say that the \av\ $z_s$ is associated with 
$[{\cal A},{\cal B}]$ or denote it by $z_s:[{\cal A},{\cal B}]$. We illustrate 
the concept of a \emph{partition} in Figure~\ref{fg.set_partition}. 

\noindent
A few partitions that play a key role in our transformation are referred to as 
forward, backward, and intermediate partitions. 

\begin{figure}[!htbp]
\begin{center}
\mbox{
\subfigure[]{\psfig{figure=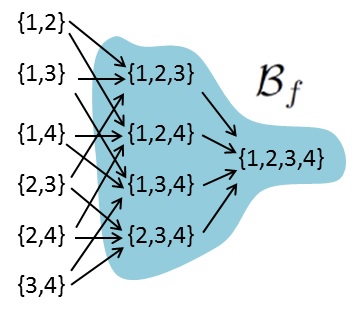,width=0.25\columnwidth}}
\subfigure[]{\psfig{figure=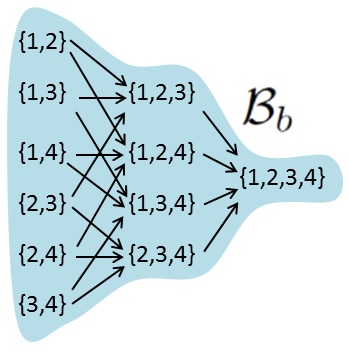,width=0.25\columnwidth}}
}
\end{center}
\caption{\it The two reference partitions, referred to as forward and backward, are shown.}
\label{fg.forward_backward}
\end{figure}

\begin{figure*}[!htbp]
\begin{center}
\mbox{
\subfigure[]{\psfig{figure=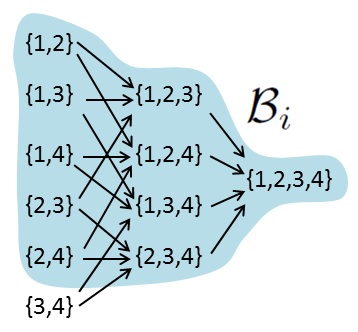,width=0.25\columnwidth}}
\subfigure[]{\psfig{figure=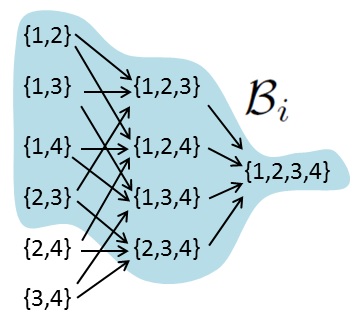,width=0.25\columnwidth}}
}
\end{center}
\caption{\it We have a total of 18 intermediate partitions. In (a), we show one of the 6 intermediate partitions 
where five sets in ${\cal B}_i$ have cardinality 2. We denote this as ${\cal I}(34)$, where the index refers 
to the only set that does not have cardinality 2. In (b), we show one of the 12 intermediate partitions where 
four sets in ${\cal B}_i$ have cardinality 2. We denote this as ${\cal I}(24,34)$, where the indices refer to 
the sets that do not have cardinality 2.
}
\label{fg.intermediate_partitions}
\end{figure*}

\begin{definition}
The {\bf forward reference partition} $[{\cal A}_f,{\cal B}_f]$ takes the form:
\begin{equation}
B \in {\cal B}_f \iff |B| \geq 3,{\cal A}_f={\cal P}\backslash{{\cal B}_f}
\end{equation}
The {\bf backward reference partition} $[{\cal A}_b,{\cal B}_b]$ is shown below:
\begin{equation}
B \in {\cal B}_b \iff |B| \geq 2,{\cal A}_b={\cal P}\backslash{{\cal B}_b}
\end{equation}

\noindent
Figure~\ref{fg.forward_backward}(a) and (b) show the forward and backward partitions respectively.
 
We consider a set of 18 partitions as {\bf intermediate partitions} $[{\cal A}_i,{\cal B}_i]$ as 
shown in Figure~\ref{fg.intermediate_partitions}. There are 6 intermediate partitions where there 
are five sets in ${\cal B}_i$ that have cardinality 2 (one such partition is shown in 
Figure~\ref{fg.intermediate_partitions}(a)). There are 12 intermediate partitions where there are 
four sets in ${\cal B}_i$ that have cardinality 2 (one such partition is shown in 
Figure~\ref{fg.intermediate_partitions}(b)). One may expect more intermediate partitions by considering 
all possible different sets in ${\cal B}_i$ having cardinality 2. However, we will see later that 
such partitions are not necessary for transforming a function in ${\cal F}_2^4$ to a function in 
${\cal F}^2$. 
\label{def.forward_backward_intermediate}
\end{definition}

\noindent
The basic idea in our work is to replace several \av s using the minimum number of 
\av s without changing the values of the function at their respective minima.

\begin{definition}
We say that a function $h({\bf x},{\bf z})$ can be transformed to another 
function $h'({\bf x},{\bf z'})$ where ${\bf z}\neq {\bf z'}$ if the 
following condition is satisfied:
\begin{equation}
\min_{\bf z}h({\bf x},{\bf z})=\min_{\bf z'}h'({\bf x},{\bf z'}),~~~~~\forall {\bf x}
\label{eq.transCondition}
\end{equation}
where ${\bf z}$ and ${\bf z'}$ are vectors of auxiliary variables with different 
partitions. The cardinality of ${\bf z}$ need not be equal to the cardinality of ${\bf z'}$.
\end{definition}

\noindent
Through a sequence of transformations of the above form, we start with a general function 
$h({\bf x},{\bf z})$ and finally compute a function $h'({\bf x},z_s,z_t)$ with only 
two \av s in reference partitions.

\begin{lemma}
Let $z_a:[{\cal A}_s,{\cal B}_s]$ and $z_b:[{\cal A}_s,{\cal B}_s]$ be two \av s that have 
the same partition then $h({\bf x},z_a,z_b) \in {\cal F}^2$ can be 
transformed to some function $h'({\bf x},z_s) \in {\cal F}^2$ involving only one \av $z_s$.
\label{lm.remove_redundant_partitions}
\end{lemma}
\begin{proof}

\noindent
According to the Equation~\ref{eq.transCondition}, we can transform a function 
$h({\bf x},z_a,z_b)$ to $h'({\bf x},z_s)$ if it satisfies the following condition:

\begin{equation}
\min_{z_a,z_b}h({\bf x},z_a,z_b)=\min_{z_s}h'({\bf x},z_s),~~~~~\forall {\bf x}
\end{equation}

\noindent
Since the \av s $z_a$ and $z_b$ take the same partition $[{\cal A}_s,{\cal B}_s]$ their 
Boolean values are equal for different configurations of ${\bf x}$. Thus we can replace 
all instances of $z_a$ and $z_b$ with $z_s$.
\qed 
\end{proof}

\begin{table}[!htbp]
\centering
\begin{tabular}{|c|c|c|c|}
\hline
Group ${\cal G}_i$ & $|{\cal G}_i|$ & 
$f({\bf x})$ &
$\min_{z_1,z_2}h({\bf x},z_1,z_2)$ where $h \in {\cal F}^2$ \\
\hline
\hline
${\cal G}_1(i,j)$ & 6 & $-x_i x_j$ & $-x_i x_j$ \\
\hline 
${\cal G}_2(i,j,k)$ & 4 & $-x_i x_j x_k$ & $\min_{z_f}(2-x_i-x_j-x_k)z_f$\\
\hline 
${\cal G}_3$ & 1 & $-x_ix_jx_kx_l$ & $\min_{z_f}(3-x_i-x_j-x_k-x_l)z_f$\\
\hline 
${\cal G}_4$ & 1 &
\begin{tabular}{c}
$-x_ix_jx_kx_l+x_ix_jx_k+x_ix_jx_l+x_ix_kx_l+$\\
$x_jx_kx_l-x_ix_j-x_ix_k-x_ix_l-$\\
$x_jx_k-x_jx_l-x_kx_l$ 
\end{tabular} &
$\min_{z_b}(1-x_i-x_j-x_k-x_l)z_b$
\\
\hline 
${\cal G}_5(i,j,k)$ & 4 &
\begin{tabular}{c}
$x_ix_jx_kx_l-x_ix_jx_k-x_ix_l-x_jx_l-$\\
$x_kx_l$ 
\end{tabular} &
$\min_{z_b}(2-x_i-x_j-x_k-2x_l)z_b$\\
\hline 
${\cal G}_6(i,j,k)$ & 4 & 
$x_ix_jx_k-x_ix_j-x_ix_k-x_jx_k$ & 
$\min_{z_b}(1-x_i-x_j-x_k)z_b$
\\
\hline 
${\cal G}_7(i)$ & 4 &
$x_ix_jx_kx_l-x_ix_jx_k-x_ix_jx_l-x_ix_kx_l$ &
$\min_{z_f}(3-2x_i-x_j-x_k-x_l)z_f$\\
\hline 
${\cal G}_8$ & 1 &
\begin{tabular}{c}
$2x_ix_jx_kx_l-x_ix_jx_k-x_ix_jx_l-x_ix_kx_l-$\\
$x_jx_kx_l$ 
\end{tabular} &
$\min_{z_f}(2-x_i-x_j-x_k-x_l)z_f$\\
\hline 
${\cal G}_9(i,j)$ & 6 &
$x_ix_jx_kx_l-x_ix_j-x_ix_kx_l-x_jx_kx_l$ &
\begin{tabular}{r}
$\min_{z_f,z_i}((2-x_k-x_l)z_f+$ \\
							 $(1-x_i-x_j)z_i$ \\
							 $- z_fz_i)$
\end{tabular} \\
\hline 
${\cal G}_{10}$ & 6 &
\begin{tabular}{c}
$-x_ix_jx_kx_l+x_ix_kx_l+x_jx_kx_l-$\\
$x_ix_k-x_ix_l-x_jx_k-x_jx_l-x_kx_l$
\end{tabular} & $f({\bf x}) \notin {\cal F}^4_2$ as shown in \cite{zivny09b}\\
\hline
\end{tabular}
\caption{\it The above table is adapted from Figure 2 of \cite{zivny08Report} where $\{i,j,k,l\}=S_4$. Each group ${\cal G}_i$ has several terms depending on the values of $\{i,j,k,l\}$. The number of distinct terms in each group is given by $|{\cal G}_i|$. Since the groups ${\cal G}_4$ and ${\cal G}_8$ involve all four variables and are symmetric, they contain one function each. $z_f$ and $z_b$ correspond to \av s for forward and backward partitions. $z_i$ corresponds to one of the intermediate partitions denoted by ${\cal I}(kl)$ in Figure~\ref{fg.intermediate_partitions}(a). For each group ${\cal G}_i$, we also use an index $(.)$ in the first column to identify a specific function from others in its group.}
\label{tb.union_of_Gs}
\end{table}

\begin{theorem}
Any function $f({\bf x})$ in ${\cal F}^4_2$ can be transformed to some function $h({\bf x},z_{f},z_s)$ in ${\cal F}^2$ where $z_f$ correspond to the forward partition and $z_s$ can either be the backward partition or one of the 18 intermediate partitions.
\label{th.reduction_to_joint}
\end{theorem}
\begin{proof}

Using Theorem 5.2 from \cite{promislow05} we can decompose a given submodular function in ${\cal F}^4$ into functions 
in 10 different groups ${\cal G}_i,i=\{1..10\}$ where each ${\cal G}_i$ is shown in Table~\ref{tb.union_of_Gs}. As 
shown in \cite{zivny09b} the  functions in ${\cal G}_{10}$ does not belong to ${\cal F}^4_2$. It was also shown that 
any submodular function that has any functions from group ${\cal G}_{10}$ does not belong to ${\cal F}^4_2$ according 
to Theorem 16(3) in \cite{zivny09b}. Thus all the functions in ${\cal F}^k_2$ should be composed of functions in the 
groups ${\cal G}_i,i \in \{1,...,9\}$. 

The number of distinct terms in each group ${\cal G}_i$ is given in Table~\ref{tb.union_of_Gs}. Overall, there are 31 
distinct functions in the groups ${\cal G}_i,i \in \{1,...,9\}$. The terms in the first group ${\cal G}_1$ has only 
second degree terms. Hence, the functions in this group does not require any \av s. The terms in the next 7 groups 
${\cal G}_i,i \in \{2,...,8\}$ can each be represented by a single \av, which can be either $z_f$ or $z_b$. Here 
$z_f$ and $z_b$ denote \av s in the forward and backward partitions respectively. The 6 terms in ${\cal G}_9$ can 
be represented using two \av s $z_f$ and $z_i$, where $z_f$ and $z_i$ correspond to forward and intermediate 
reference partitions (denoted by ${\cal I}(k,l)$ in Figure~\ref{fg.intermediate_partitions}(a)) respectively. It is 
important to note that the functions in ${\cal G}_9$ involve interaction between $z_f$ and $z_i$, i.e., there exists 
a bilinear term $z_fz_i$ in ${\cal G}_9$. 

We prove the result by considering two cases. 

\paragraph{Absence of ${\cal G}_9$ functions:}
In the first case, we consider functions that can be expressed as a sum of 
functions in the first 8 groups ${\cal G}_i,i=\{1..8\}$. In other words, we study the scenario where we express the 
function without using any function from ${\cal G}_9$. Let us denote such a function as $f0({\bf x})$ that can be expressed 
as a sum of 25 functions from the 8 groups ${\cal G}_i,i=\{1..8\}$ as shown below:

\begin{equation}
f0({\bf x})  			   = \alpha_{1} {\cal G}_1(i,j)   + 
											 \cdots                        
											 \alpha_{25}{\cal G}_8          
\label{eq.sumofgroups_case0}
\end{equation}

The only \av s involved in all the functions are $z_f$ and $z_b$. Using Lemma~\ref{lm.remove_redundant_partitions} we can 
obtain a function that uses only two variables $z_f$ and $z_b$ as shown below:

\begin{equation}
f0({\bf x}) =  g({\bf x}) + \min_{z_f,z_b} ( g_f({\bf x})z_f + g_b({\bf x})z_b),
\label{eq.sumofgroups_case0_h}
\end{equation}
where $g({\bf x})$, $g_f({\bf x})$ and $g_b({\bf x})$ are functions involving ${\bf x}$. This implies that any function that can 
be expressed without any function from ${\cal G}_9$ can be expressed using the only forward and backward partitions. 

\paragraph{Presence of ${\cal G}_9$ functions:}

Let us consider an arbitrary function $f({\bf x})$ in ${\cal F}^4_2$ that is expressed as a sum of 
functions from these 31 groups including functions from ${\cal G}_9$:

\begin{equation}
f({\bf x})  =  \alpha_1 {\cal G}_1(i,j) + \alpha_2 {\cal G}_2(i,j) + \cdots + \alpha_{31} {\cal G}_9(k,l)
\label{eq.sumofgroups}
\end{equation}

In Table~\ref{tb.pairwiseSums}, we show that the sum of two functions can always be 
represented using only two auxiliary variables. In Tables~\ref{tb.tripleSums},~\ref{tb.quadrupleSums} and \ref{tb.fiveSums} we show 
the sum of functions with 3, 4 and 5 terms respectively. Different combinations of functions lead to functions that can always be 
expressed with only 2 auxiliary variables.

Without loss of generality, we have avoided the repetition for all possible indices by treating them using the 
set $\{i,j,k,l\} = \{1,2,3,4\}$. We have already proved the case where where ${\cal G}_9$ is absent. 
Thus, the tables only show summations that involve at least one function from ${\cal G}_9$. 

In some cases, we do not show the sums of functions with real coefficients $(\alpha,\beta,\gamma,\delta,\eta)$ to demonstrate the 
special scenarios where the combination of two functions involving intermediate partition $z_i$ can be transformed to a function that 
involves only $z_f$ and $z_b$. In such cases, we can do the following sequentially:

\begin{equation}
\alpha_1 f_1({\bf x}) + \alpha_2 f_2({\bf x}) = 
\alpha_1 (f_1({\bf x}) + f_2({\bf x})) + (\alpha_2 - \alpha_1) f_2({\bf x}),~~~~~\alpha_1 \le \alpha_2
\end{equation}

Let $\beta = \alpha_2 - \alpha_1$ and let $f_3({\bf x}) = f_1({\bf x}) + f_2({\bf x})$ as per the Table~\ref{tb.pairwiseSums}. Now we can further use Table~\ref{tb.pairwiseSums} on $\alpha_1 f_3({\bf x}) + \beta f_2({\bf x})$ to generate other functions. 

As we see in Table~\ref{tb.union_of_Gs}, the function ${\cal G}_9(i,j)$ uses two auxiliary variables $z_f$ and $z_i \in {\cal I}(k,l)$. As we observe in Table~\ref{tb.pairwiseSums}, on adding the function ${\cal G}_9(i,j)$ with other functions we have the following  scenarios:

\begin{enumerate}
\item The coefficient of $z_i$ is unaltered and we only change the coefficients of $z_f$. This happens in 6 of the additions in Table~\ref{tb.pairwiseSums} as given by $({\cal G}_9(i,j), {\cal G}_1(i,j))$, $({\cal G}_9(i,j),{\cal G}_2(j,k,l))$, $({\cal G}_9(i,j),{\cal G}_7(k))$ and $({\cal G}_9(i,j),{\cal G}_8)$. 

\item We obtain a function that can be expressed with only one $z_f$. This happens in 4 of the additions in Table~\ref{tb.pairwiseSums} as given by $({\cal G}_9(i,j),{\cal G}_2(i,j,k))$, $({\cal G}_9(i,j),{\cal G}_3)$,$({\cal G}_9(i,j),{\cal G}_7(i))$, and $({\cal G}_9(i,j),{\cal G}_9(k,l))$.

\item We obtain a function that can be expressed with only one $z_b$. This happens in 2 of the additions shown in Table~\ref{tb.pairwiseSums} as given by $({\cal G}_9(i,j),{\cal G}_4)$ and $({\cal G}_9(i,j),{\cal G}_6(j,k,l))$. 

\item We obtain a function that can be expressed with $z_f$ and $z_b$. This happens in one of the additions shown in Table~\ref{tb.pairwiseSums} as given by $({\cal G}_9(i,j),{\cal G}_5(i,j,k))$.
   
\item We obtain a function with $z_f$ and $z_i$, whose coefficients are changed. This happens in 3 of the additions shown in Table~\ref{tb.pairwiseSums} as given by $({\cal G}_9(i,j),{\cal G}_5(j,k,l))$, $({\cal G}_9(i,j),{\cal G}_6(i,j,k))$, and $({\cal G}_9(i,j),{\cal G}_9(i,k))$.  
\end{enumerate} 

There are 6 functions in group ${\cal G}_9$. However, as shown in Table~\ref{tb.pairwiseSums}, additions involving ${\cal G}_9(i,j)$ and ${\cal G}_9(k,l)$ produce functions involving only one auxiliary variable $z_f$. In other words, 
sum of functions involving ${\cal G}_9(i,k)$ can sometimes be represented using functions from the first 8 groups (${\cal G}_1$ to ${\cal G}_8$). Out of the 6 functions in ${\cal G}_9$ only two of them are necessary at a time. Without loss of generality, we rewrite the $f({\bf x})$ using a maximum of 2 functions in group ${\cal G}_9$ as shown below:

\begin{equation}
f({\bf x})  			   =  \alpha_{1} {\cal G}_1(i,j) +
												\cdots +
												\alpha_{26} {\cal G}_9(i,j) + 
											  \alpha_{27} {\cal G}_9(i,k). 
\label{eq.sumofgroups_case_removeG9s}
\end{equation}

\noindent
The remaining four terms in ${\cal G}_9$ are not necessary due to the following reasons:

\begin{itemize}
\item ${\cal G}_9(i,l)$ is not necessary because its addition to ${\cal G}_9(i,j)$ and ${\cal G}_9(i,k)$ will lead to a function involving only $z_f$ and $z_b$ as per the second last row of Table~\ref{tb.tripleSums}.
\item Any function ${\cal G}_9(j,k)$ is not necessary because its addition to ${\cal G}_9(i,j)$ and ${\cal G}_9(i,k)$ will lead to a function involving only $z_f$ and $z_b$ as per the last row of Table~\ref{tb.tripleSums}.
\item Any function ${\cal G}_9(j,l)$ is not necessary because its addition to ${\cal G}_9(i,k)$ can be represented using a function that involves only $z_f$ as per the last row of Table~\ref{tb.pairwiseSums}.
\item Any function ${\cal G}_9(k,l)$ is not necessary because its addition to ${\cal G}_9(i,j)$ can be represented using a function that involves only $z_f$ as per the last row of Table~\ref{tb.pairwiseSums}.
\end{itemize}

\noindent
We observed that we need a maximum of two functions from ${\cal G}_9$ to represent any function in ${\cal F}^4_2$. So there are 
two possibilities for $f({\bf x})$ and we denote them as $f1({\bf x})$ and $f2({\bf x})$ depending on whether we use one or two 
of the functions from ${\cal G}_9$ as shown below:

\begin{align}
f1({\bf x})  =  & \alpha {\cal G}_9(i,j) + 
	  				     \beta {\cal G}_5(i,k,l) + 
					       \gamma {\cal G}_5(j,k,l) + 
							   \delta {\cal G}_6(i,j,k) + 
							   \eta {\cal G}_6(i,j,l) + 
								 \sigma \min_{z_f} (g_f({\bf x})z_f) + g({\bf x}), \\ 
f2({\bf x})  =  & \alpha {\cal G}_9(i,j) + 
	  				      \beta  {\cal G}_9(i,k) + 
					        \gamma {\cal G}_5(j,k,l) + 
							    \delta {\cal G}_6(i,j,k) + 
							    \eta \min_{z_f} (g_f({\bf x})z_f) + g({\bf x}).
\label{eq.sumofgroups_case1and2}
\end{align}

We have represented $f1$ and $f2$ using 7 and 6 terms respectively. We show that we do not need any other functions for representing $f1$ and $f2$:

\begin{itemize}

\item ${\cal G}_1$: We can represent them using $g({\bf x})$. 

\item ${\cal G}_2$: These functions involve $z_f$ and we can represent them using $\min_{z_f} g_f({\bf x},z_f)$ according to Lemma~\ref{lm.remove_redundant_partitions}. 

\item ${\cal G}_3$: These functions involve $z_f$ and we can represent them using $\min_{z_f} g_f({\bf x},z_f)$ according to Lemma~\ref{lm.remove_redundant_partitions}. 

\item ${\cal G}_4$: By adding this function to ${\cal G}_9(i,j)$ we obtain functions in ${\cal G}_6$. The generated ${\cal G}_6$ functions can be subsequently added to any functions in ${\cal G}_9(i,j)$ or ${\cal G}_9(i,k)$. If there is no ${\cal G}_9$ term, then we can represent the function using $z_f$ and $z_b$ as explained earlier in the case of $f0$. 

\item ${\cal G}_5:$ In the case of $f1$, some functions in ${\cal G}_5$ can be added to ${\cal G}_9(i,j)$ to obtain ${\cal G}_6$ and ${\cal G}_8$. The generated functions can be subsequently added to any ${\cal G}_9(i,j)$ terms. If there is no ${\cal G}_9(i,j)$ term, then we can represent the function using $z_f$ and $z_b$ as explained earlier in the case of $f0$. The functions ${\cal G}_5(i,k,l)$ and ${\cal G}_5(j,k,l)$ alter the coefficients of $z_i$. For now, we keep these functions of ${\cal G}_5$ as separate terms in function $f1({\bf x})$. 

\noindent
In the case of $f2$, some functions in ${\cal G}_5$ can be added to ${\cal G}_9(i,j)$ or ${\cal G}_9(i,k)$ to obtain ${\cal G}_6$ and ${\cal G}_8$ terms. The generated functions can be subsequently added to any ${\cal G}_9(i,j)$ or ${\cal G}_9(i,k)$ terms. If there is no ${\cal G}_9(i,j)$ term, then we can represent the function using $z_f$ and $z_b$ as explained earlier in the case of $f0$. On adding the function ${\cal G}_5(j,k,l)$ to ${\cal G}_9(i,j)$ or ${\cal G}_9(i,k)$ we produce functions that alter the coefficients of $z_i$. For now, we keep this function in ${\cal G}_5$ to represent the original function $f2({\bf x})$.

\item ${\cal G}_6:$ In the case of $f1$, some functions in ${\cal G}_6$ can be added to ${\cal G}_9(i,j)$ to obtain functions in ${\cal G}_5$, which can be subsequently added to any ${\cal G}_9(i,j)$ terms. If there is no ${\cal G}_9(i,j)$ term, then we can represent the function using $z_f$ and $z_b$ as explained earlier in the case of $f0$. The functions ${\cal G}_6(i,j,k)$ and ${\cal G}_6(i,j,l)$ produce functions that alter the coefficients of $z_i$. For now, we keep these two functions of ${\cal G}_6$ as separate terms in $f1({\bf x})$.

\noindent
In the case of $f2$, some functions in ${\cal G}_6$ can be added to ${\cal G}_9(i,j)$ or ${\cal G}_9(i,k$ to obtain ${\cal G}_5$, which can be subsequently added to any ${\cal G}_9(i,j)$ or ${\cal G}_9(i,k)$ terms. If there is no ${\cal G}_9(i,j)$ term, then we can represent the function using $z_f$ and $z_b$ as explained earlier in the case of $f0$. On adding the function ${\cal G}_6(i,j,k)$ to ${\cal G}_9(i,j)$ or ${\cal G}_9(i,k)$ we produce functions that alter the coefficients of $z_i$. For now, we keep this function ${\cal G}_6(i,j,k)$ as separate term in $f2({\bf x})$.

\item ${\cal G}_7:$ Some functions in ${\cal G}_7$ can be added to ${\cal G}_9(i,j)$ or ${\cal G}_9(i,k)$ to generate functions in ${\cal G}_8$, which can be subsequently added to any ${\cal G}_9(i,j)$ or ${\cal G}_9(i,k)$ terms. In other cases, the functions in ${\cal G}_7$ only modify the coefficients of $z_f$ terms that can be represented by the function $\min_{z_f}(g_f({\bf x})z_f)$. 

\item ${\cal G}_8:$ This function can be represented by $\min_{z_f}(g_f({\bf x})z_f)$ since it only has one \av $z_f$.

\end{itemize}

\begin{table}[!htbp]
\centering
\begin{tabular}{|c|c|r|c|}
\hline
$f_1({\bf x}),\alpha \in {\mathbb R}^{+}$ & 
$f_2({\bf x}),\beta  \in {\mathbb R}^{+}$ &  
\begin{tabular}{c}
$\min_{z_1,z_2}h({\bf x},z_1,z_2),$ \\ 
where 
$h({\bf x},z_1,z_2) = f_1({\bf x}) + f_2({\bf x}), ~~~~~\forall {\bf x}$ 
\end{tabular} & 
\av s
\\
\hline
\hline
$\alpha {\cal G}_9(i,j)$ & 
$\beta {\cal G}_1(i,j)$ &
\begin{tabular}{r}
$-\beta x_ix_j + \alpha \min_{z_f,z_i}((2-x_k-x_l)z_f + $ \\
$(1-x_i-x_j)z_i - $ \\
$z_fz_i)$
\end{tabular} &
\begin{tabular}{c}
$z_f$,\\
$z_i \in {\cal I}(kl)$
\end{tabular} 
\\
\hline
${\cal G}_9(i,j)$ & 
${\cal G}_2(i,j,k)$ &
\begin{tabular}{r}
$-x_ix_j + {\cal G}_7(k) =$ \\
$-x_ix_j + \min_{z_f}(3 - x_i - x_j - 2x_k - x_l)z_f$
\end{tabular} &
$z_f$
\\
\hline
$\alpha {\cal G}_9(i,j)$ & 
$\beta {\cal G}_2(j,k,l)$ &
\begin{tabular}{r}
$\min_{z_f,z_i}(( \alpha (2-x_k-x_l) + \beta (2 - x_j - x_k - x_l) )z_f + $ \\
							  $ \alpha (1-x_i-x_j)z_i - $ \\
							  $ \alpha z_fz_i)$
\end{tabular} &
\begin{tabular}{c}
$z_f$,\\
$z_i \in {\cal I}(kl)$
\end{tabular} 
\\
\hline
${\cal G}_9(i,j)$ & 
${\cal G}_3$ &
\begin{tabular}{r}
$-x_ix_j + {\cal G}_2(i,k,l) + {\cal G}_2(j,k,l) =$\\
$-x_ix_j + \min_{z_f}((2 - x_i - x_k - x_l) + (2 - x_j - x_k - x_l))z_f$
\end{tabular} &
$z_f$
\\
\hline
${\cal G}_9(i,j)$ & 
${\cal G}_4$ &
\begin{tabular}{r}
${\cal G}_6(i,j,k) + {\cal G}_6(i,j,l)$ - $x_kx_l$= \\
$-x_kx_l$ + $\min_{z_b}(2-2x_i-2x_j-x_k-x_l)z_b$
\end{tabular} &
$z_b$ \\
\hline
${\cal G}_9(i,j)$ & 
${\cal G}_5(i,j,k)$ &
\begin{tabular}{r}
${\cal G}_8 + {\cal G}_6(i,j,l)$ - $x_kx_l$= \\
$-x_kx_l$ + $\min_{z_f}(2-x_i-x_j-x_k-x_l)z_f + $ \\
$\min_{z_b}(1-x_i-x_j-x_l)z_b$
\end{tabular} &
\begin{tabular}{c}
$z_f$,\\
$z_b$
\end{tabular} 
\\
\hline
$\alpha {\cal G}_9(i,j)$ & 
$\beta {\cal G}_5(j,k,l)$ &
\begin{tabular}{r}
$\min_{z_f,z_i}(( \alpha (2-x_k-x_l)z_f +$\\
							 $(\alpha (1 - x_i - x_j) + \beta (2-2x_i-x_j-x_k-x_l))z_i - $ \\
							 $ \alpha z_fz_i)$
\end{tabular} &
\begin{tabular}{c}
$z_f$,\\
$z_i \in {\cal I}(kl)$
\end{tabular} 
\\
\hline
$\alpha {\cal G}_9(i,j)$ & 
$\beta {\cal G}_6(i,j,k)$ &
\begin{tabular}{r}
$\min_{z_f,z_i}((\alpha (2-x_k-x_l)z_f +$ \\
							 $(\alpha (1-x_i-x_j) + \beta (1-x_i-x_j-x_k))z_i - $ \\
							 $ \alpha z_fz_i)$
\end{tabular} &
\begin{tabular}{c}
$z_f$,\\
$z_i \in {\cal I}(kl)$
\end{tabular} 
\\
\hline
${\cal G}_9(i,j)$ & 
${\cal G}_6(j,k,l)$ &
\begin{tabular}{r}
${\cal G}_5(i,k,l) - x_k x_l = $ \\
$-x_kx_l + \min_{z_b}(2-x_i-2x_j-x_k-x_l)z_b$
\end{tabular} &
$z_b$ \\
\hline
${\cal G}_9(i,j)$ & 
${\cal G}_7(i)$ &
\begin{tabular}{r}
${\cal G}_8 - x_ix_j + {\cal G}_2(i,k,l) =$ \\
$-x_1x_2 + \min_{z_f}((2-x_i-x_j-x_k-x_l) + (2-x_i-x_k-x_l))z_f$
\end{tabular} &
$z_f$\\
\hline
$\alpha {\cal G}_9(i,j)$ & 
$\beta {\cal G}_7(k)$ &
\begin{tabular}{r}
$\min_{z_f,z_i}((\alpha (2-x_k-x_l) + \beta (3 - x_i - x_j - 2x_k - x_l))z_f +$ \\
							 $\alpha (1-x_i-x_j)z_i -$ \\
							 $\alpha z_fz_i)$
\end{tabular} &
\begin{tabular}{c}
$z_f$,\\
$z_i \in {\cal I}(kl)$
\end{tabular} 
\\
\hline
$\alpha {\cal G}_9(i,j)$ & 
$\beta {\cal G}_8$ &
\begin{tabular}{r}
$\min_{z_f,z_i}(\alpha (2 - x_k - x_l) + \beta (2-x_i-x_j-x_k-x_l))z_f + $ \\
							 $\alpha (1-x_i-x_j)z_i - $ \\
							 $\alpha z_fz_i)$
\end{tabular} &
\begin{tabular}{c}
$z_f$,\\
$z_i \in {\cal I}(kl)$
\end{tabular} 
\\
\hline
$\alpha {\cal G}_9(i,j)$ & 
$\beta {\cal G}_9(i,k)$ &
\begin{tabular}{r}
$\min_{z_f,z_i}((\alpha (2-x_k-x_l) + \beta (2 - x_j - x_l))z_f + $ \\
							 $(\alpha (1-x_i-x_j) + \beta (1 - x_i - x_k))z_i$ -\\
							 $\alpha z_fz_i - \beta z_fz_i)$
\end{tabular} &
\begin{tabular}{c}
$z_f$,\\
$z_i \in {\cal I}(kl,jl)$
\end{tabular} 
\\
\hline
${\cal G}_9(i,j)$ & 
${\cal G}_9(k,l)$ &
\begin{tabular}{r}
${\cal G}_8 - x_ix_j - x_kx_l =$ \\
$-x_ix_j - x_kx_l + \min_{z_f}((2-x_i-x_j-x_k-x_l)z_f$ 
\end{tabular} &
$z_f$ 
\\
\hline
\end{tabular}
\caption{\it We show the sum of a function ${\cal G}_9(i,j)$ with any other function in Table~\ref{tb.union_of_Gs} can be expressed 
using two auxiliary variables. Here the index set $\{i,j,k,l\}=S_4$ denotes the four distinct integers $S_4 = \{1,2,3,4\}$.}
\label{tb.pairwiseSums}
\end{table}

\begin{table}[!htbp]
\centering
\begin{tabular}{|c|c|c|r|c|}
\hline
$f_1({\bf x}),\alpha \in {\mathbb R}^{+}$ & 
$f_2({\bf x}),\beta \in {\mathbb R}^{+}$ &  
$f_3({\bf x}),\gamma \in {\mathbb R}^{+}$ &  
\begin{tabular}{r}
$\min_{z_1,z_2}h({\bf x},z_1,z_2),$ \\ 
$h({\bf x},z_1,z_2) = f_1({\bf x}) + f_2({\bf x}) + f_3({\bf x}), ~~~~~\forall {\bf x}$ 
\end{tabular} & 
\av s
\\
\hline
\hline
$\alpha {\cal G}_9(i,j)$ &
$\beta {\cal G}_5(i,k,l)$ &
$\gamma {\cal G}_5(j,k,l)$ &
\begin{tabular}{r}
$\min_{z_f,z_i}((\alpha (2-x_k-x_l)z_f + $ \\
							 $(\alpha (1-x_i-x_j)+$ \\
							 $\beta (2 - x_i - 2x_j - x_k - x_l) + $ \\
							 $\gamma (2 - 2x_i - x_j -x_k - x_l))z_i - $ \\							  
							 $\alpha z_fz_i)$
\end{tabular} &
\begin{tabular}{r}
$z_f$,\\
$z_i \in {\cal I}(kl)$
\end{tabular} 
\\
\hline
$\alpha {\cal G}_9(i,j)$ &
$\beta {\cal G}_6(i,j,k)$ &
$\gamma {\cal G}_6(i,j,l)$ &
\begin{tabular}{r}
$\min_{z_f,z_i}((\alpha (2-x_k-x_l)z_f$ \\
							 $(\alpha (1-x_i-x_j) + $ \\
							 $ \beta (1-x_i-x_j -x_k) + $ \\
							 $ \gamma (1-x_i-x_j - x_l))z_i - $ \\
							 $ \alpha z_fz_i)$
\end{tabular} &
\begin{tabular}{r}
$z_f$,\\
$z_i \in {\cal I}(kl)$
\end{tabular} 
\\
\hline
$\alpha {\cal G}_9(i,j)$ &
$\beta {\cal G}_5(j,k,l)$ &
$\gamma {\cal G}_6(i,j,l)$ &
\begin{tabular}{r}
$\min_{z_f,z_i}((\alpha (2-x_k-x_l)z_f$ \\
							 $(\alpha (1-x_i-x_j) + $ \\
							 $ \beta (2-2x_i-x_j -x_k -x_l) + $ \\
							 $ \gamma (1-x_i-x_j - x_l))z_i - $ \\
							 $ \alpha z_fz_i)$
\end{tabular} &
\begin{tabular}{r}
$z_f$,\\
$z_i \in {\cal I}(kl)$
\end{tabular} 
\\
\hline
$\alpha {\cal G}_9(i,j)$ &
$\beta {\cal G}_9(i,k)$ &
$\gamma {\cal G}_5(j,k,l)$ &
\begin{tabular}{r}
$\min_{z_f,z_i}((\alpha (2-x_k-x_l) + \beta (2 - x_j - x_l))z_f$ \\
							 $(\alpha (1-x_i-x_j) + $ \\
							 $ \beta  (1-x_i-x_k) + $ \\
							 $ \gamma (2-2x_i-x_j -x_k- x_l))z_i - $ \\
							 $ \alpha z_fz_i - \beta z_fz_i)$
\end{tabular} &
\begin{tabular}{r}
$z_f$,\\
$z_i \in {\cal I}(kl,jl)$
\end{tabular} 
\\
\hline
$\alpha {\cal G}_9(i,j)$ &
$\beta {\cal G}_9(i,k)$ &
$\gamma {\cal G}_6(i,j,k)$ &
\begin{tabular}{r}
$\min_{z_f,z_i}((\alpha (2-x_k-x_l) + \beta (2 - x_j - x_l))z_f$ \\
							 $(\alpha (1-x_i-x_j) + $ \\
							 $ \beta  (1-x_i-x_k) + $ \\
							 $ \gamma (1-x_i-x_j -x_k))z_i - $ \\
							 $ \alpha z_fz_i - \beta z_fz_i)$
\end{tabular} &
\begin{tabular}{r}
$z_f$,\\
$z_i \in {\cal I}(kl,jl)$
\end{tabular} 
\\
\hline
${\cal G}_9(i,j)$ &
${\cal G}_9(i,k)$ &
${\cal G}_9(i,l)$ &
\begin{tabular}{r}
$ - x_jx_kx_l + {\cal G}_8 + {\cal G}_5(j,k,l) = $ \\
$-x_jx_kx_l + \min_{z_f}(2 - x_i - x_j - x_k - x_l)z_f + $ \\ 
$\min_{z_b}(2 - 2x_i - x_j - x_k - x_l)z_b$
\end{tabular} & 
\begin{tabular}{r}
$z_f$, \\
$z_b$
\end{tabular}
\\
\hline
${\cal G}_9(i,j)$ &
${\cal G}_9(i,k)$ &
${\cal G}_9(j,k)$ &
\begin{tabular}{r}
${\cal G}_7(l) + {\cal G}_8 + {\cal G}_6(i,j,k) = $ \\
$\min_{z_f}((3 - x_i - x_j - x_k - 4x_l) + $\\
$(2 - x_i - x_j - x_k - x_l))z_f + $ \\ 
$\min_{z_b}(1 - x_i - x_j - x_k )z_b$
\end{tabular} & 
\begin{tabular}{r}
$z_f$, \\
$z_b$
\end{tabular}
\\
\hline
\end{tabular}
\caption{\it We show the sum of any three functions from Table~\ref{tb.union_of_Gs} can be expressed 
using two auxiliary variables. Here the index set $\{i,j,k,l\}=S_4$ denotes the four distinct 
integers $S_4 = \{1,2,3,4\}$.}
\label{tb.tripleSums}
\end{table}

\begin{table}[!htbp]
\centering
\begin{tabular}{|c|c|c|c|r|c|}
\hline
$f_1({\bf x})$ & 
$f_2({\bf x})$ &  
$f_3({\bf x})$ &  
$f_4({\bf x})$ &  
\begin{tabular}{r}
$\min_{z_1,z_2}h({\bf x},z_1,z_2),$ \\ 
$h({\bf x},z_1,z_2) = f_1({\bf x}) + f_2({\bf x}) + f_3({\bf x})+ f_4({\bf x}), ~~~~~\forall {\bf x}$ 
\end{tabular} & 
\av s
\\
\hline
\hline
${\cal G}_9(i,j)$ &
${\cal G}_5(i,k,l)$ &
${\cal G}_5(j,k,l)$ &
${\cal G}_5(i,j,k)$ &
\begin{tabular}{r}
$({\cal G}_9(i,j) + {\cal G}_5(i,j,k)) + $\\
$({\cal G}_5(i,k,l) + {\cal G}_5(j,k,l)) =$\\
$({\cal G}_8 + {\cal G}_6(i,j,l) + {\cal G}_1(k,l)) + $\\
$({\cal G}_5(i,k,l) + {\cal G}_5(j,k,l))$\\
\end{tabular} &
\begin{tabular}{r}
$z_f$,\\
$z_b$
\end{tabular} 
\\
\hline
${\cal G}_9(i,j)$ &
${\cal G}_6(i,j,k)$ &
${\cal G}_6(i,j,l)$ &
${\cal G}_6(j,k,l)$ &
\begin{tabular}{r}
$({\cal G}_9(i,j) + {\cal G}_6(j,k,l)) +$\\
$({\cal G}_6(i,j,k) + {\cal G}_6(i,j,l)) =$\\
$({\cal G}_5(i,j,k) + {\cal G}_1(k,l)) + $\\
$({\cal G}_6(i,j,k) + {\cal G}_6(i,j,l))$\\
\end{tabular} &
\begin{tabular}{r}
$z_f$,\\
$z_b$
\end{tabular} 
\\
\hline
\hline
$\alpha {\cal G}_9(i,j)$   &
$\beta {\cal G}_5(i,k,l)$  &
$\gamma {\cal G}_5(j,k,l)$ &
$\delta {\cal G}_6(i,j,k)$ &
\begin{tabular}{r}
$\min_{z_f,z_i}(\alpha (2-x_k-x_l)z_f + $ \\
							 $(\alpha (1 - x_i - x_j) + $\\
							 $ \beta  (2 - x_i - 2x_j - x_k - x_l)+$\\
							 $ \gamma (2 - 2x_i - x_j - x_k - x_l)+$\\
							 $ \delta (1 - x_i - x_j - x_k))z_i - $\\
							 $ \alpha z_fz_i)$ 
\end{tabular} &
\begin{tabular}{r}
$z_f$,\\
$z_i \in {\cal I}(k,l)$
\end{tabular} 
\\
\hline
$\alpha {\cal G}_9(i,j)$ &
$\beta {\cal G}_6(i,j,k)$ &
$\gamma {\cal G}_6(i,j,l)$ &
$\delta {\cal G}_5(i,k,l)$ &
\begin{tabular}{r}
$ \min_{z_f,z_i}(\alpha (2 - x_k - x_l)z_f + $ \\
							 $(\alpha (1 - x_i - x_j) + $\\
							 $ \beta  (1 - x_i - x_j - x_k)+$\\
							 $ \gamma (1 - x_i - x_j - x_l)+$\\
							 $ \delta (2 - x_i - 2x_j - x_k -x_l))z_i - $\\
							 $ \alpha z_fz_i)$ 
\end{tabular} &
\begin{tabular}{r}
$z_f$,\\
$z_i \in {\cal I}(k,l)$
\end{tabular} 
\\
\hline
$\alpha {\cal G}_9(i,j)$ &
$\beta {\cal G}_9(i,k)$ &
$\gamma {\cal G}_5(j,k,l)$ &
$\delta {\cal G}_6(i,j,k)$ &
\begin{tabular}{r}
$ \min_{z_f,z_i}((\alpha (2 - x_k - x_l) + \beta (2 - x_j - x_l))z_f + $ \\
							 $(\alpha (1 - x_i - x_j) +$\\ 
							 $ \beta  (1 - x_i - x_k) +$\\
							 $ \gamma (2 - 2x_i - x_j - x_k - x_l)+$\\
							 $ \delta (1 - x_i - x_j - x_k))z_i - $\\
							 $ \alpha z_fz_i - \beta z_fz_i)$ 
\end{tabular} &
\begin{tabular}{r}
$z_f$,\\
$z_i \in {\cal I}(k,l)$
\end{tabular} 
\\
\hline
\end{tabular}
\caption{\it We show the sum of any four functions from Table~\ref{tb.union_of_Gs} can be expressed 
using two auxiliary variables. Here the index set $\{i,j,k,l\}=S_4$ denotes the four distinct 
integers $S_4 = \{1,2,3,4\}$.}
\label{tb.quadrupleSums}
\end{table}

\begin{table}[!htbp]
\centering
\begin{tabular}{|c|c|c|c|c|c|}
\hline
\begin{tabular}{r}
$f_1({\bf x}),$\\
$\alpha \in {\mathbb R}^{+}$ 
\end{tabular} & 
\begin{tabular}{r}
$f_2({\bf x}),$\\
$\alpha \in {\mathbb R}^{+}$ 
\end{tabular} & 
\begin{tabular}{r}
$f_3({\bf x}),$\\
$\alpha \in {\mathbb R}^{+}$ 
\end{tabular} & 
\begin{tabular}{r}
$f_4({\bf x}),$\\
$\alpha \in {\mathbb R}^{+}$ 
\end{tabular} & 
\begin{tabular}{r}
$f_5({\bf x}),$\\
$\alpha \in {\mathbb R}^{+}$ 
\end{tabular} & 
\begin{tabular}{r}
$\min_{z_1,z_2}(f_1({\bf x}) + f_2({\bf x}) + f_3({\bf x}) + f_4({\bf x}) + f_5({\bf x}))$,\\
$~~~~~\forall {\bf x}$ 
\end{tabular}
\\
\hline
\hline
$\alpha {\cal G}_9(i,j)$ &
$\beta {\cal G}_5(i,k,l)$ &
$\gamma {\cal G}_5(j,k,l)$ &
$\delta {\cal G}_6(i,j,k)$ &
$\eta {\cal G}_6(i,j,l)$ &
\begin{tabular}{r}
$\min_{z_f,z_i}(\alpha (2-x_k-x_l)z_f +$ \\
$(\alpha (1 - x_i - x_j)+$\\
$ \beta  (2 - x_i - 2x_j - x_k - x_l)+$\\
$ \gamma (2 -2x_i -  x_j - x_k - x_l)+$\\
$ \delta (1 - x_i - x_j - x_k)+$\\
$ \eta   (1 - x_i - x_j - x_l))z_i-$\\
$ \alpha z_fz_i)$ 
\end{tabular} 
\\
\hline
\end{tabular}
\caption{\it We show the sum of five functions from Table~\ref{tb.union_of_Gs} can be expressed using two auxiliary 
variables $z_f$ and $z_i \in {\cal I}(k,l)$. Here the index set $\{i,j,k,l\}=S_4$ denotes the four distinct 
integers $S_4 = \{1,2,3,4\}$.}
\label{tb.fiveSums}
\end{table}

Using Table~\ref{tb.fiveSums} we rewrite $f1$ as given below:

\begin{align}
f1({\bf x})  = & g({\bf x}) + \min_{z_f,z_i}((\alpha (2-x_k-x_l) + \sigma g({\bf x}))z_f + \\ \nonumber
							 &	 (\alpha (1 - x_i - x_j)+ \\ \nonumber
							 &		\beta  (2 - x_i - 2x_j - x_k - x_l)+ \\ \nonumber
							 &		\gamma (2 -2x_i -  x_j - x_k - x_l)+ \\ \nonumber
							 &		\delta (1 - x_i - x_j - x_k)+ \\ \nonumber
							 &		\eta   (1 - x_i - x_j - x_l))z_i- \\ \nonumber
							 &    \alpha z_f z_i) \\ \nonumber
						 = &g({\bf x}) + \min_{z_f,z_i} (g'_f({\bf x})z_f + g_i({\bf x})z_i - \alpha z_f z_i) \\
\end{align}
									
Using the last row of Table~\ref{tb.quadrupleSums} we rewrite $f2$ as given below:
								
\begin{align}
f2({\bf x}) = & g({\bf x}) + \min_{z_f,z_i}((\alpha (2 - x_k - x_l) + \beta (2 - x_j - x_l) + g_f({\bf x}))z_f +  \\ \nonumber
					    & (\alpha (1 - x_i - x_j) +\\ \nonumber 
						  &  \beta  (1 - x_i - x_k) +\\ \nonumber
						  &  \gamma (2 - 2x_i - x_j - x_k - x_l)+\\ \nonumber
						  &  \delta (1 - x_i - x_j - x_k))z_i - \\ \nonumber
						  &  \alpha z_fz_i - \beta z_fz_i) \\ \nonumber
					 =  & g({\bf x}) + \min_{z_f,z_i} (g'_f({\bf x})z_f + g_i({\bf x})z_i - (\alpha+\beta) z_f z_i) \\
\end{align}
						
It is shown that $f1$ and $f2$ need only two \av s $z_f$ and $z_s$. In the case of $f0$, $z_s$ is a backward partition. In the case of $f1$ and $f2$, $z_s$ belongs to one of the 18 intermediate partitions.
\qed
\end{proof}

\section{Linear Programming solution}
\label{sec.algorithm}
For a given function $f(x_1,x_2,x_3,x_4)$ in ${\cal F}^4_2$, our goal is to compute 
a function $h({\bf x},{\bf z})$ in ${\cal F}^2$. Theorem~\ref{th.reduction_to_joint} 
shows that we need only two \av s $(z_{f},z_{s})$. Here $z_f$ corresponds to the 
forward reference partition. The \av $z_s$ is either the backward partition or one 
of the 18 intermediate reference partitions. Unfortunately, we do not know which 
one of these 19 partitions is required before we do the transformation. In what 
follows, we will show the transformation assuming that we know the specific 
partition for $z_s$. Note that $z_b$ is a special case of $z_i$ and we do not use 
the bilinear term $z_fz_s$ when $z_s = z_b$. In order to handle this condition we use 
a Boolean variable that takes the value $0$ when the intermediate partition is the 
backward reference partition and $1$ otherwise:

\begin{equation}
\delta(z_s) = \begin{cases} 0 & \text{if $z_i \in [{\cal A}_b,{\cal B}_b],$}\\
1 & \text{otherwise.} 
\end{cases}
\label{eq.eta}
\end{equation} 

\noindent
The required function $h({\bf x},{\bf z})$ is the following:
\begin{equation}
h({\bf x},z_{f},z_{s}) = b_0 + 
												\sum_{i}b_ix_i - 
												\sum_{i>j} b_{ij}x_ix_j + 
												(g_{f}-\sum_{i=1}^4g_{f,i}x_i)z_{f} +
												(g_{s}-\sum_{i=1}^4g_{s,i}x_i)z_{s} - 
												\delta(z_s) j_{fs}z_{f}z_{s},
\label{eq.function_hat}
\end{equation}
such that $b_{ij},g_{f,i},g_{s,i},j_{fs}\geq 0$ and $i,j\in S_4$. As we know the partitions of 
$(z_{f},z_{s})$, we know their Boolean values for all labelings of ${\bf x}$. We need the 
coefficients \mbox{$(b_i,b_{ij},j_{fs},g_{f},g_{s},g_{f,i},g_{s,i}),i \in S_4$} to compute 
$h(x_1,x_2,x_3,x_4,z_{f},z_{s})$. These coefficients satisfy both submodularity constraints 
(that the coefficients of all bilinear terms \mbox{$(x_ix_j,x_iz_{f},x_{j}z_{s},z_{f}z_{s})$} 
are less than or equal to zero) and those imposed by the reference partitions. First we list 
the submodularity conditions below:
\begin{equation}
\underbrace{
\begin{pmatrix} 
b_{ij} \\
g_{f,i} \\
g_{s,i} \\
j_{fs}
\end{pmatrix}^T}_{{\cal S}_p} 
\geq 
\mathbf{0},i,j=S_4,i \neq j,
\label{eq.submodularity}
\end{equation}
where $\mathbf{0}$ refers of a vector composed of $0$'s of appropriate 
length. Next we list the conditions which guarantee \mbox{$f({\bf
    x})=\min_{z_{f},z_{s}}h({\bf x},z_{f},z_{s})$}, $~~~~~\forall {\bf x}$. 
Let $\eta(S)$ be the value of $z_{f}z_{s}$ for $S \in {\cal P}$. This 
can be obtained using the partitions of $z_f$ and $z_s$.

\begin{equation}
\eta(S) = \begin{cases} 1 & \text{if $S \in ({\cal B}_f \cap {\cal B}_s)$}\\
0 & \text{otherwise.} 
\end{cases}
\label{eq.eta}
\end{equation} 

\noindent
Let us denote the value of \av $z_s$ for different subsets of $S_4$ as given 
below:
\begin{equation}
z_s^{S} = \begin{cases} 1 & \text{if $S \in {\cal B}_s,$}\\
0 & \text{if $S \in {\cal A}_s.$} 
\end{cases}
\label{eq.eta}
\end{equation} 

\noindent
Let ${\cal G}$ and ${\cal H}$ denote values of functions $f$ and $h$ respectively:

\begin{equation}
{\cal G}=f(\mathbf{1}^S_1,\mathbf{1}^S_2,\mathbf{1}^S_3,\mathbf{1}^S_4),~~~~~\forall S \in {\cal P}
\end{equation}

\begin{equation}
{\cal H}=h(\mathbf{1}^S_1,\mathbf{1}^S_2,\mathbf{1}^S_3,\mathbf{1}^S_4,0,0) +
			   (g_{f}-\sum_{i=1}^4g_{f,i}\mathbf{1}^S_i)z_f^S + 
				 (g_{s}-\sum_{i=1}^4g_{s,i}\mathbf{1}^S_i)z_s^S -
				 \delta(z_s) \eta(S) j_{fs} 
\end{equation}

\noindent
As a result we have the following 16 linear Equations (N.B. there are $2^4(16)$ different $S$):

\begin{equation}
{\cal G}={\cal H},~~~~~\forall S \in {\cal P}
\label{eq.trans_equality}
\end{equation}

\noindent
We already know the partition of $(z_{f},z_{s})$ and their appropriate values a priori. The following constraints ensure that $z_{f}$ and $z_{s}$ behave according to their associated partitions. 

\begin{eqnarray}
\underbrace{
\begin{pmatrix}
g_f-\sum_{i=1}^4g_{f,i}\mathbf{1}^S_i\\
g_s-\sum_{i=1}^4g_{s,i}\mathbf{1}^D_i\\
\end{pmatrix}}_{{\cal G}_g}
&\geq&
\mathbf{0},S \in {\cal A}_{f},D \in {\cal A}_{s} 
\\
\underbrace{
\begin{pmatrix}
g_f-\sum_{i=1}^4g_{f,i}\mathbf{1}^S_i - \delta (z_s) \eta(S) j_{fs}\\
g_s-\sum_{i=1}^4g_{s,i}\mathbf{1}^D_i - \delta (z_s) \eta(D) j_{fs}\\
\end{pmatrix}}_{{\cal G}_l}
 &\leq&
\mathbf{0},S \in {\cal B}_{f},D \in {\cal B}_{s}.
\label{eq.greaterlessthan}
\end{eqnarray}

\noindent
We need to compute the coefficients
$(b_{ij},g_{f},g_{f,i},g_{s},g_{s,i},j_{fs})$ that satisfy the 
Equations~(\ref{eq.submodularity}),~(\ref{eq.trans_equality}), and (\ref{eq.greaterlessthan}). This is equivalent to finding a feasible point in a linear programming problem:

\begin{equation}
\min~const
\end{equation}
\begin{equation}
s.t~{\cal S}_p \geq \mathbf{0},~{\cal G}={\cal H},~{\cal G}_g\geq \mathbf{0},~{\cal
  G}_l\leq \mathbf{0}
\label{eq.lp_constraints}
\end{equation}

\noindent
In the above LP formulation we assumed that we know the partition of \av s $z_f$ 
and $z_s$. However, $z_s$ can be one of the 19 partitions. Before we do the transformation 
it is not easy to know which one of the 19 partitions is necessary. So we solve the LP 19 times 
and iterate over all the 19 partitions to identify the necessary one. For the correct 
partition, will be able to find a solution that satisfies all the constraints.

\section{Experiments}

The functions in the class ${\cal F}^4_2$ can be transformed to functions in ${\cal F}^2$ using 25 
\av s according to existing results \cite{zivny09a}. We show that this transformation can be done 
using only two \av s using a linear program. In Matlab, the transformation takes around 0.03 seconds 
and it can be further improved using efficient C++ implementation. 

\begin{figure}[!htbp]\centering
{\psfig{figure=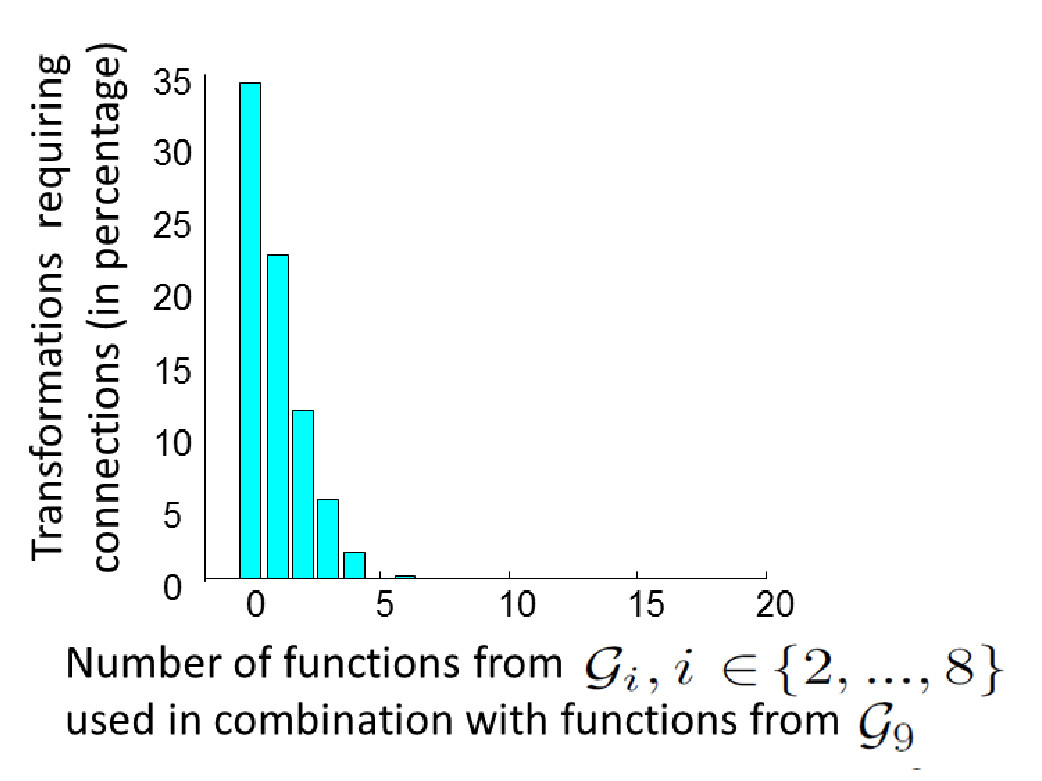,width=0.50\columnwidth}}
\caption{\it We generated submodular functions using non-weighted sum of functions from groups ${\cal G}_i,i=\{2,...,9\}$. The x-axis denotes the number of functions chosen 
from ${\cal G}_i,i=\{2,...,8\}$ in generating the submodular functions and the y-axis gives the percentage of transformations requiring $z_i$ in intermediate partition. 
}
\label{fg.Experiments}
\end{figure}


In our experiments as shown in Figure~\ref{fg.Experiments}, we generated submodular functions using non-weighted sum of functions from 
${\cal G}_9$ and functions from groups ${\cal G}_i,i=\{2,...,8\}$. We do not consider functions 
from group ${\cal G}_1$ since they do not require any \av s. The number of functions $n_{\cal G}$ used 
from groups ${\cal G}_i , i=\{2,...,8\}$ is increased from 0 to 19. For each value of $n_{\cal G}$, 
we generated 1000 functions and the non-negative weights are randomly generated in the interval $[0,1]$. 
We observed that as we increase $n_{\cal G}$, the generated submodular functions were less likely to 
use intermediate \av s. This also concurs with Table~\ref{tb.pairwiseSums}, that many combinations 
of ${\cal G}_9$ with other functions can be represented using functions in the first 
8 groups (${\cal G}_1$ to ${\cal G}_8$) that do not require any \av s in intermediate partition.

\section{Discussion and open problems}
The reduction of higher order functions to quadratic ones will be beneficial for developing efficient minimization algorithms. These techniques can be broadly classified into two types: submodularity-preserving 
\cite{billionnet85,kolPAMI04,hammer65,zalesky03,queyranne95,FreedmanD05,rhys1970,zivny09b,Rother2009} and 
general techniques \cite{rosenberg80,fix11,gallagher11,boros2012,ishikawa2011}. This paper belongs to the 
submodular-preserving class of algorithms where higher order submodular functions are transformed to quadratic submodular functions using \av s. The general techniques are usually employed in association with roof-duality approaches for minimizing non-submodular functions \cite{boros1989,BorosH02,boros2008,rother07}. The general techniques also employ \av s and these \av s need not be \mbf s. The existing upper bound for general reduction techniques is given by $G(k)=2^{k-2}(k-3)+1$ for a $k$th order function. We show the comparison between the \av s used in general techniques and submodularity-preserving techniques in Table~\ref{tb.DedvsIshikawa}. 

\begin{table}[!htbp]
\centering
\begin{tabular}{cccccccc}
\hline
Type & Degree & 3 & 4 & 5 & 6 & 7 & 8 \\
\hline
General & Ishikawa \cite{ishikawa2011} & 1 & 5 & 17 & 49 & 129 & 321 \\
\hline
Submodularity Preserving & Dedekind \cite{kleitman69} & 1 & 2 & 7581 & $\approx7.8 \times 10^6$ & $\approx 2.4\times 
10 ^{12}$ & $\approx 5.6\times 10^{23}$ \\
\hline
\end{tabular}
\caption{Comparison of the number of \av s used for general versus submodularity-preserving techiques. The Dedekind number $D(k)$ is unknown for $k>8$.}
\label{tb.DedvsIshikawa}
\end{table}

Note that the upper bound for the number of \av s required for submodularity-preserving transformation is much higher than for general reduction techniques. We have improved the upper bound for submodular functions from $2^{2^k}$ to Dedekind number $D(k)$. In the case of fourth order functions we have further improved the upper bound from 168 ($D(4)$) to 2. 
\newline

\noindent
{\bf Acknowledgments:} Srikumar Ramalingam would like to thank Mitsubishi Electric Research Laboratories (MERL) for the support. \v Lubor Ladick\'{y} is funded by Max Planck Center for Learning Systems Fellowship. Philip H.S. Torr would like to acknowledge the financial support provided by ERC grant ERC-2012-AdG 321162-HELIOS, EPSRC/MURI grant ref EP/N019474/1, EPSRC grant EP/M013774/, and EPSRC Programme Grant Seebibyte EP/M013774/1.

{\small
\bibliographystyle{plain}
\bibliography{dam}
}

\end{document}